%% file: main.tex
\documentclass[10pt,twocolumn,letterpaper]{article}

\usepackage[pagenumbers]{wacv} 

\input{preamble}

\definecolor{wacvblue}{rgb}{0.21,0.49,0.74}
\usepackage[pagebackref,breaklinks,colorlinks,allcolors=wacvblue]{hyperref}

\def\wacvPaperID{2604} 
\def\confName{WACV}
\def\confYear{2026}
\usepackage{graphicx}
\usepackage{subcaption}
\usepackage[accsupp]{axessibility}

\title{Curve Skeletonization in Continuous domain for Meshes and Point Clouds}

\author{Jai Bardhan \\
TCS Research\\
{\tt\small jai.bardhan@cvut.cz}
\and
Ramya Hebbalaguppe\\
TCS Research\\
{\tt\small ramya.hebbalaguppe@tcs.com}
\and
Aravind Udupa\\
IIT Delhi\\
{\tt\small ara.udupa@gmail.com}
}

\begin{document}

\twocolumn[{%
\renewcommand\twocolumn[1][]{#1}%
\maketitle
\begin{center}
    \centering
    \captionsetup{type=figure}
    \includegraphics*[width=\textwidth]{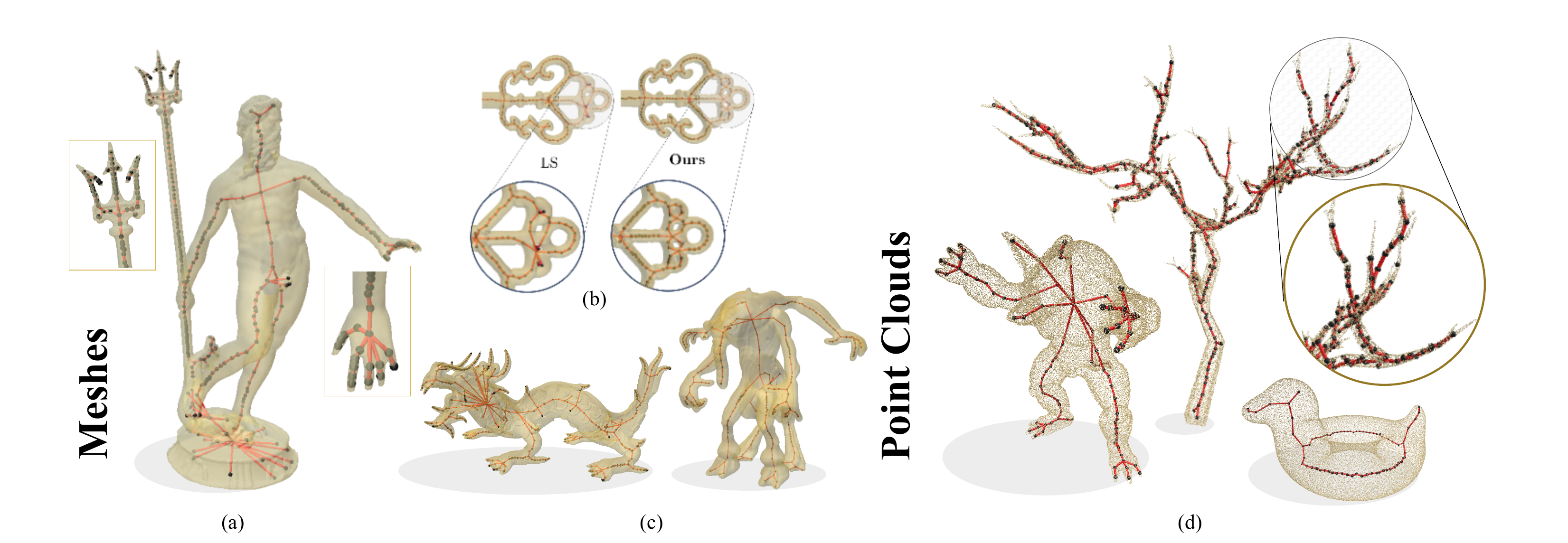}
    \captionof{figure}{\textbf{Representative results of the proposed \cscd on diverse 3D shapes from various benchmark datasets - Meshes (left) and Point Clouds (right):} (a) We show the result of our method on the \texttt{neptune} mesh -- The inset illustrates the excellent skeletal quality for both the hand and the trident; (b) We show a comparison of \cscd to a contemporary method (LS~\cite{origLS}) for \texttt{Copper-key}  -- \cscd reconstructs the holes of the shape better; (c) \cscd captures fine details in complex meshes like \texttt{xyzrgb-dragon} (Stanford Library) and \texttt{TID:133568} (Thingi10k); (d) Our framework generalizes across domains, performing well on point clouds such as \texttt{bob}, \texttt{armadillo}, and the intricate \texttt{dead tree}. See Appendix~\ref{sec:append-results} for more results.
    }
    \label{fig:main_figure}
\end{center}%
}]

\input{sections/0_abstract}    
\input{sections/1_intro}

\input{sections/pre3_CSCD}
\input{sections/new3_proposedMethod}
\input{sections/4_results}

\input{sections/6_conclusions}
{
    \small
    \bibliographystyle{ieeenat_fullname}
    \bibliography{sample-base}
}

\input{sections/supplementary}

\end{document}

%% file: preamble.tex
\usepackage{xspace} 
\usepackage[dvipsnames]{xcolor} 
\usepackage{pifont} 
\usepackage{wrapfig} 
\usepackage{algorithm}      
\usepackage{algorithmicx}   
\usepackage{algpseudocode}  
\usepackage{titletoc} 
\usepackage{subcaption}

\colorlet{mypink}{red!40}
\colorlet{myblue}{cyan!60}

\def\cscd{\texttt{CSCD}\xspace}

\newcommand{\xmark}{\ding{55}}

\algrenewcommand\algorithmicrequire{\textbf{Input:}}
\algrenewcommand\algorithmicensure{\textbf{Output:}}

\usepackage[dvipsnames]{xcolor}

\def\cscd{\texttt{CSCD}\xspace}

\usepackage[dvipsnames]{xcolor}
\usepackage{pifont}

%% file: sections/0_abstract.tex
\begin{abstract}

Advancements in 3D curve skeletonization are accelerating progress across a wide range of applications. However, developing robust skeletonization algorithms that capture intricate object details remains challenging. Skeletonization via Local Separators (LS) offers an efficient graph-based approach but suffers from representation inaccuracies due to its discrete nature. To address this, we introduce CSCD, a novel framework for Curve Skeletonization in the Continuous Domain, generalizing LS to manifolds. Specifically, we present two realizations: CSCD-M for meshes and CSCD-PC for point clouds. CSCD-M leverages the intrinsic triangulation of a mesh for resilience to noise and improved topological preservation, while CSCD-PC employs tufted Laplacians for enhanced robustness. To our knowledge, CSCD-M is the first intrinsic method for curve skeletonization. Our results show CSCD-M matches LS performance across diverse meshes and outperforms LS (TOG'21) on benchmarks like Thingi10k dataset. CSCD-PC qualitatively outperforms CoverageAxis++ (Eurographics'24) and EPCS (CAG'23). Finally, we demonstrate the efficacy of CSCD in a few downstream tasks: object classification, shape segmentation, identifying handles, tunnels, and constrictions in objects. 

\noindent\textbf{Project Website:} \url{https://cscd-skel.pages.dev}

\end{abstract}

%% file: sections/1_intro.tex
\section{Introduction}
\label{sec:intro}

3D object representation is a fundamental problem in computer graphics/
vision, as it aims to capture the shape, structure, and appearance of objects in a digital format. Various representations have been developed over the years, each with strengths and limitations. Meshes~\cite{mesh_proessing, deep_learning_mesh}, point clouds, distance fields~\cite{distance_fields}, and recently NeRFs~\cite{nerf, nerf_review} are commonly used for geometry processing. These representations, however, can be  detailed or complex for certain applications, especially in shape and motion modeling. 

Curve skeletons have emerged as a powerful alternative. They capture the topology and geometry of the object through a set of 1-D connected curves that lie in the medial axis of the object and approximate key geometrical features. Curve skeletons are invaluable for a multitude of applications including shape segmentation~\cite{skel_report_2}, matching~\cite{GOH2008326}, retrieval~\cite{skel_shape_match}, animation~\cite{transfer4d}, reconstruction~\cite{skel_recon}.

\noindent \textbf{Challenges in curve skeletonization:}
Despite their utility, the computation of curve skeletons is fraught with challenges, necessitating algorithms that are both robust and sensitive to nuances such as capturing fine-grained shapes and structure~\cite{skel_report, skel_report_2}. There is a lack of a clear definition of curve skeletons for 3D objects. This has led to a multitude of hand-crafted methods, each with its strengths and weaknesses. Most of these methods rely on the idea that for tubular shapes, there exists a 1D structure that preserves the shape of the topology. Among them: (1) geometric features-based methods rely on identifying key geometric features of the shape, but struggle with complex shapes or noisy data; (2) local decimation methods progressively simplify the objects while maintaining the topological structure, but generally fail to capture the high fidelity features; (3) division based methods divide the shape into regions, and compute skeleton points for these regions; (4) learning based methods like~\cite{Lin2020Point2SkeletonLS, p2mat} use machine learning techniques to generate skeletons, but usually fail to generalize to unseen objects, and (5) Medial Axis Transform based methods identify the medial axis/plane and prune to obtain the curve skeletons. Representative works include:
\cite{Au_mesh_contraction,tag_mc,Livesu2012ReconstructingTC,cheng_dual,cov_axis, genus_0,nicu2007curve, cornea2005curve, LI2023209, wang2024coverage, dou2022coverage, chu2023robustly,li2015q,wang2022computing}. 
Recently, a skeletonization technique based on local separators (LS) \cite{origLS} has shown particular promise to produce curve skeletons with higher fidelity, capturing the finer details of the shape, where methods such as MCF~\cite{tag_mc} and $L_1$-medial skeletonization~\cite{huang} sometimes fall short. The LS method works by constructing local separators on graphs to divide the graph locally into non-overlapping regions, then calculating the centroid for each region to form the nodes of the skeleton. We summarize the native domain (representation) of operation for a few methods in Tab.~\ref{tab:rep-operation}. We want to take note that while point cloud/graph based methods may be applied to meshes, they are not natively developed for meshes and therefore miss out face-level information available.

Although the LS method achieves strong results, it is limited by its discrete representation, which can produce noisy curve skeletons and sensitivity to input quality (e.g., poorly triangulated meshes or noisy point clouds). Moreover, the absence of a continuous formulation hinders its applicability to continuous manifolds and restricts integration with representation-specific algorithms (Sec.~\ref{sec:ours_vs_ls}).

\begin{figure}
    \centering
    \includegraphics[width=0.7\linewidth]{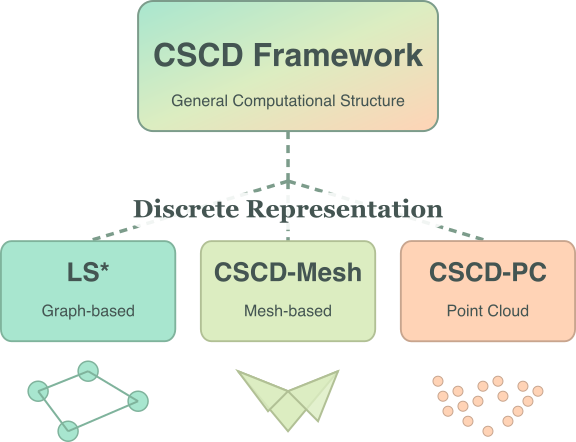}
    \caption{\textbf{CSCD Framework Overview:} Our framework generalizes LS~\cite{origLS}, such that a graph-based realization results in an algorithm similar to LS (Appendix~\ref{sec:append-ls-cscd}), mesh-based realization leads to \cscd-M, and a point cloud realization leads to \cscd-PC (Sec.~\ref{sec:realizations}).}
    \label{fig:cscd-framework}
\end{figure}

\begin{table}[t]
    \centering
    \resizebox{0.45\textwidth}{!}{
    \begin{tabular}{c|c|c|c}
    \toprule
        Method & Graph & Mesh & Point Cloud \\
        \toprule
        LS~\cite{origLS} & \textcolor{ForestGreen}{\ding{51}} & \textcolor{BrickRed}{\xmark} & \textcolor{BrickRed}{\xmark} \\
        MSLS~\cite{fastLS} & \textcolor{ForestGreen}{\ding{51}} & \textcolor{BrickRed}{\xmark} & \textcolor{BrickRed}{\xmark} \\
        ROSA~\cite{rosa} & \textcolor{BrickRed}{\xmark} & \textcolor{BrickRed}{\xmark} & \textcolor{ForestGreen}{\ding{51}} \\
        MCF~\cite{tag_mc} & \textcolor{BrickRed}{\xmark} & \textcolor{ForestGreen}{\ding{51}} & \textcolor{BrickRed}{\xmark} \\
        EPCS~\cite{LI2023209} & \textcolor{BrickRed}{\xmark} & \textcolor{BrickRed}{\xmark} & \textcolor{ForestGreen}{\ding{51}} \\
        CA++~\cite{dou2022coverage,wang2024coverage} & \textcolor{BrickRed}{\xmark} & \textcolor{ForestGreen}{\ding{51}} & \textcolor{ForestGreen}{\ding{51}} \\
        \midrule
        \textbf{\cscd (Ours)} & \textcolor{ForestGreen}{\ding{51}} (App.~\ref{sec:append-ls-cscd}) & \textcolor{ForestGreen}{\ding{51}}\cscd-M & \textcolor{ForestGreen}{\ding{51}}\cscd-PC \\
        \bottomrule
    \end{tabular}
    }
    \caption{Native representations for various curve skeletonization algorithms. \cscd enables construction of skeletonization algorithms tailored to each representation.}
    \label{tab:rep-operation}
\end{table}

\begin{figure*}[t]
    \small \centering
    \includegraphics[width=\textwidth]{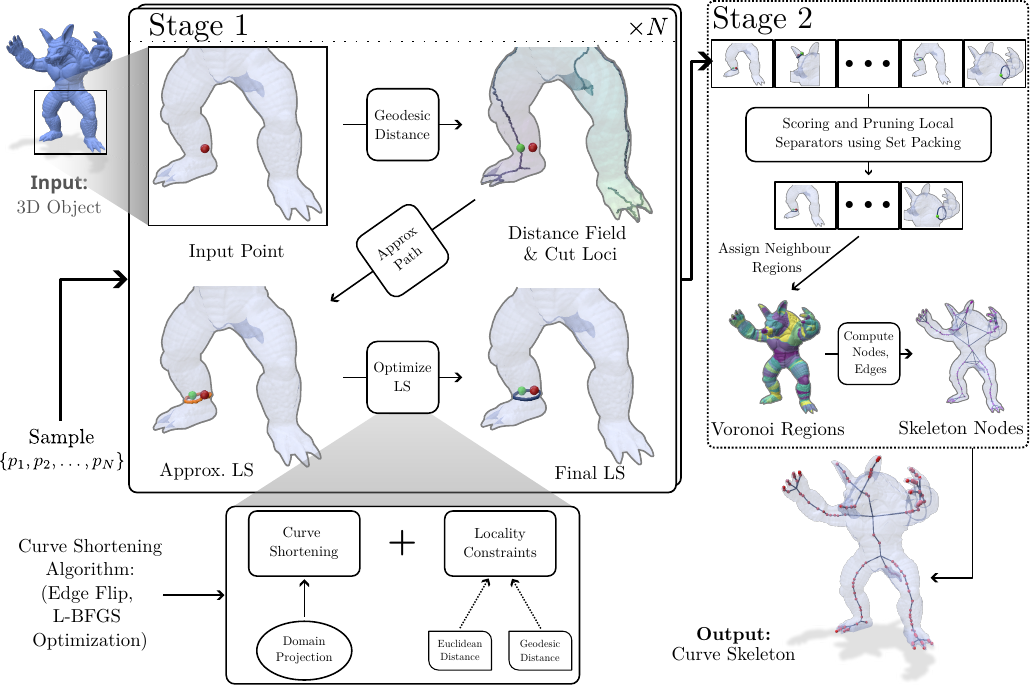}
    \caption{
        \textbf{[Schematic of \cscd for Curve Skeletonization].} The entire Curve Skeletonization framework can be divided into two stages. In Stage 1, we calculate the various local separators given the 3D object as input(input can be a mesh or a point cloud). Once we have a sampled point, we calculate the geodesic distance to all other points and find the cut loci of the point to identify the target cut locus. Then an approximate local separator (LS) is constructed, followed by the LS optimization. The optimization is specific to the particular representation and is in the presence of a locality constraint. 
        In Stage 2, we calculate the skeleton from the set of local separators. First, we prune and pack the previously obtained separators. Based on the obtained separators, we divide the object into Voronoi regions. These regions correspond to nodes, and neighbouring regions are connected to give the skeleton. Post-processing is finally performed to convert cliques to stars in the resultant skeleton. 
    }
    \label{fig:proc-diag}
\end{figure*}

Motivated by high-quality results in \cite{origLS} and the need to address the limitations, we introduce \cscd, a novel framework for \textbf{C}urve \textbf{S}keletonization in the \textbf{C}ontinuous \textbf{D}omain, generalizing LS to manifolds (see Fig.~\ref{fig:cscd-framework}).

\noindent \textbf{Rationale for \cscd:} \cscd operates on manifolds rather than discrete structures like graphs, offering advantages for 3D shape analysis, particularly of surface and geometric features \cite{tierny:hal-00725576,kimmel1998computing,surazhsky2005fast}. (1) As continuous representations, manifolds more accurately capture intrinsic geometry and topology (e.g., curvature, geodesic distances), whereas graph discretization can distort these properties. (2) \cscd leverages differential geometry tools such as the Laplace-Beltrami operator and intrinsic vector fields, which are robust to non-rigid deformations. By using domain specific implementations, we can reduce discretization artefacts that are present in graph-based methods approximations.

Our \textbf{key contributions} include:
\begin{enumerate}
    \item[(1)] We introduce the \cscd framework, a generalization of the LS algorithm beyond graph representations (see, Fig.~\ref{fig:cscd-framework}, Tab.~\ref{tab:rep-operation}). 
    \item[(2)] We introduce \cscd-M, a realization of \cscd for meshes that operates upon the intrinsic triangulation of mesh. This is the first method to operate on intrinsic triangulation offering robustness by construction. \cscd-M performs comparable or better than LS and is $\sim60\%$ faster, on average, on our set of meshes of varying sizes and complexity (see Tab. \ref{tab:timing-inference}).
    \item[(3)] Capitalizing on the generality of \cscd, we also introduce \cscd-PC, a realization of \cscd to point clouds.
    \item[(4)] As a crucial component of our framework, we adapt and improve upon existing cut locus identification strategies. For CSCD-M, we develop an intrinsic formulation of the algorithm, leading to improved robustness on poorly triangulated meshes. For CSCD-PC, we introduce a novel cut locus identification strategy tailored for point clouds (see. Fig.~\ref{fig:res-comparion-cut-locus-identify}, Fig.~\ref{fig:res-cl-pc} and Appendix.~\ref{sec:practical-cl-adapt}).
    \item[(5)] Finally, we demonstrate that our framework can be minimally modified to approximately identify handles, tunnels, and constricting loops, thereby extending its applicability beyond skeletonization (see App. Fig.~\ref{fig:tunnel-handle-append}). Our improved skeletonization yields better results for downstream applications, (App. Fig.~\ref{fig:shape-segmentation-append}, Tab.~\ref{tab:res-shape-classification}). See, App.~\ref{sec:append-downstream-complete}. 
\end{enumerate}

%% file: sections/pre3_CSCD.tex
\section{CSCD}
\label{sec:cscd-desc}

\subsection{\cscd Framework}

\cscd is a framework for local separators based skeletonization on manifolds.  Our method takes a shape $X \in \mathbb{R}^3$ as input, where $X$ can be any 3D representation with discrete differential operators (Eg., gradient and Laplace-Beltrami). Our goal is to generate a curve skeleton $C$ from the input.

\begin{algorithm}[t]
  \caption{\cscd\ Framework (Detailed alg. in Supplementary \ref{sec:pseudocode})}
  \label{alg:cscd-overview-alg}
  \begin{algorithmic}[1]
    \Require A 3D object \(\mathcal{O}\)
    \Ensure The curve skeleton of \(\mathcal{O}\) as \(\mathcal{C}\)

    \State \(\mathcal{P} \gets\) set of points on surface \(\mathcal{S}\)
    \For{\(p \in \mathcal{P}\)}
      \State \(D \gets\) geodesic distances \(f(p,\mathcal{S})\)
      \State \(C \gets\) cut-locus mask \(f(D,\mathcal{S})\)
      \State \(t \gets\) target cut locus from \((C,D,p,\mathcal{S})\)
      \State \(\hat l \gets\) traced path from \(t\) to \(p\)
      \State \(l \gets\) locally optimized separator from \(\hat l\)
    \EndFor

    \State \(L \gets \{l_1,\dots,l_{|\mathcal{P}|}\}\)
    \State \(M \gets\) overlap map where \(M_{i,j}=1\) if loops \(i,j\) overlap
    \State \(\mathcal{L} \gets\) separators after packing \((M,L)\)
    \State \(\mathcal{R} \gets\) nearest region assignments \(f(\mathcal{L},\mathcal{S})\)
    \State \(\hat N \gets\) node positions \(f(\mathcal{R},\mathcal{S})\)
    \State \(\hat E \gets\) edge connectivity \(f(\mathcal{R},\mathcal{S})\)
    \State \((N,E) \gets\) clique-cleaned graph from \((\hat N,\hat E)\)
    \State \(\mathcal{C}\gets(N,E)\)
  \end{algorithmic}
\end{algorithm}

Our framework comprises two stages: \textbf{(1)} Finding a set of local separators that divide the manifold locally into two halves; \textbf{(2)} scoring and selecting an optimal set of non-overlapping separators through set-packing. We then assign the nearest region of the shape to each local separator, calculate centroids to obtain skeleton nodes, and connect nodes of neighboring regions. Finally, we remove cliques to form the curve skeleton. Refer to Algorithm~\ref{alg:cscd-overview-alg} for further details, where, $f(\cdot)$ denotes functions specific to each step.

\subsection{Stage 1: Local Separator Construction} \label{sec:sep_const}

\begin{figure}
    \centering
    \includegraphics[width=0.7\linewidth]{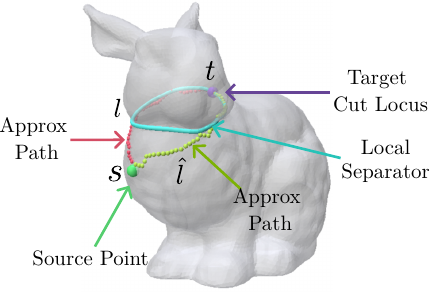}
    \caption{\textbf{[Illustration of a local separator.]} The \textcolor[rgb]{0.388, 0.922, 0.506}{\textbf{green}} vertex is the source $s$, the \textcolor[rgb]{0.667, 0.553, 0.831}{\textbf{purple}} vertex is the target cut locus $t$ (Sec.~\ref{sec:cut-locus-identify}). The two paths $\hat{l}$ (\textcolor[rgb]{0.82, 0.329, 0.408}{\textbf{red}} and \textcolor[rgb]{0.718, 0.875, 0.369}{\textbf{lime}}) are the approximate paths (Sec.~\ref{sec:trace-path}), and the \textcolor[rgb]{0.40, 0.91, 0.875}{\textbf{cyan}} curve $l$ is the final local separator (Sec.~\ref{sec:loop-optim}). See Sec.~\ref{sec:append-vocab} for Terminology/definitions. }
    \label{fig:curve-shorten}
\end{figure}

An ideal local separator is a locally short path that divides the surface into two halves with the following properties:
\begin{enumerate}
    \item[\textbf{P1:}] The separator goes around geometrical features (protrusions) rather than simply dividing the local surface.
    \item[\textbf{P2:}] The separator is \textit{locally shortest} within a defined locality.
\end{enumerate}

To construct a local separator, we start with a source point (Fig.~\ref{fig:curve-shorten}) and calculate geodesic distances. Geodesics are curves on the manifold that locally minimize distance, i.e., generalizing straight lines to curved spaces. To satisfy property \textbf{P1}, the separator passes through a cut locus of the source. \textbf{Cut loci are points on the manifold where multiple minimizing geodesics from the source intersect.} At cut loci, the gradient of the distance field is not defined and the laplacian is $+\infty$. For manifolds with boundaries, we identify separator extremities as boundary points where the sum of geodesic distance gradients cancels out, similar to cut loci. In our implementations (Sec.~\ref{sec:realizations}), we omit this case.

One can visualize a circular wave emanating from the source: when the wavefront meets a protrusion, it splits and meets at the cut locus. 
The ideal local separator is the minimal loop connecting the wavefront split point to the cut locus and back. We construct these separators by optimizing an approximate loop from the target cut locus to the source through locally constrained curve shortening.

To create a set of local separators, we sample multiple (say, $N$) source points and repeat this procedure.

\subsection{Stage 2: Constructing the Curve Skeleton}
After obtaining potentially overlapping local separators, we need to select non-overlapping ones to divide the object into Voronoi regions. We score each separator and prune them through greedy set packing~\cite{Kordalewski2013NewGH}. From the final set of non-overlapping separators, we identify neighboring regions and calculate average positions of points within regions to obtain skeleton nodes. We connect neighboring regions to form skeletal edges and remove cliques (e.g., triangles formed by connecting three neighboring regions) through iterative removal to create the final curve skeleton.

\subsection{CSCD vs. LS}
\label{sec:ours_vs_ls}
The LS procedure does not readily extend to continuous manifolds. In particular, (1) growing the separator set is non-trivial, as naively adding nearby points is inefficient and diverges from the original method, and (2) the absence of a clear neighborhood structure on manifolds complicates stopping criteria. We address these challenges by proposing a novel framework—a strict generalization of LS—for continuous manifolds.

%% file: sections/new3_proposedMethod.tex
\section{Realization of \cscd on meshes  /point clouds}
\label{sec:realizations}

Building on the above framework, we propose a method for both meshes (\cscd-M) and point clouds (\cscd-PC).
\subsection{Stage1: Constructing Local Separators}

\subsubsection{Choice of the Geodesic Distance Method}

We use the heat method for geodesic distance computation due to its efficiency and accuracy~\cite{Crane:2017:HMD}, making it well-suited for mesh and point cloud processing.

\subsubsection{Identification of the Cut Loci}
\label{sec:cut-locus-identify}
For both \cscd-M and \cscd-PC, we adapt the practical cut locus algorithm from~\cite{cutLocusComp}, as detailed in Appendix~\ref{sec:practical-cl-adapt}. The algorithm works by, starting from the farthest cut locus, identifying the cut loci as a connected graph on the surface of the mesh.

\paragraph{For \cscd-M,} we adapt the algorithm for intrinsic triangulation, ensuring robustness to poor meshing. All gradient calculations in the procedure are restricted to the tangent spaces of vertices and faces through barycentric interpolation. Fig.~\ref{fig:res-comparion-cut-locus-identify}, Fig.~\ref{fig:comparion-cut-locus-identify} (in Appendix) compares our intrinsic implementation with the original on a poorly triangulated mesh.

\paragraph{For \cscd-PC,} we introduce a novel method for cut locus identification on point clouds, with its reliability demonstrated in Fig.~\ref{fig:res-cl-pc}, Fig.~\ref{fig:cut-locus-pc} (in Appendix).

\subsubsection{Selecting the Target Cut Locus}
\label{sec:target-cl}
\begin{wrapfigure}{r}{0pt}
    \includegraphics[width=0.45\linewidth]{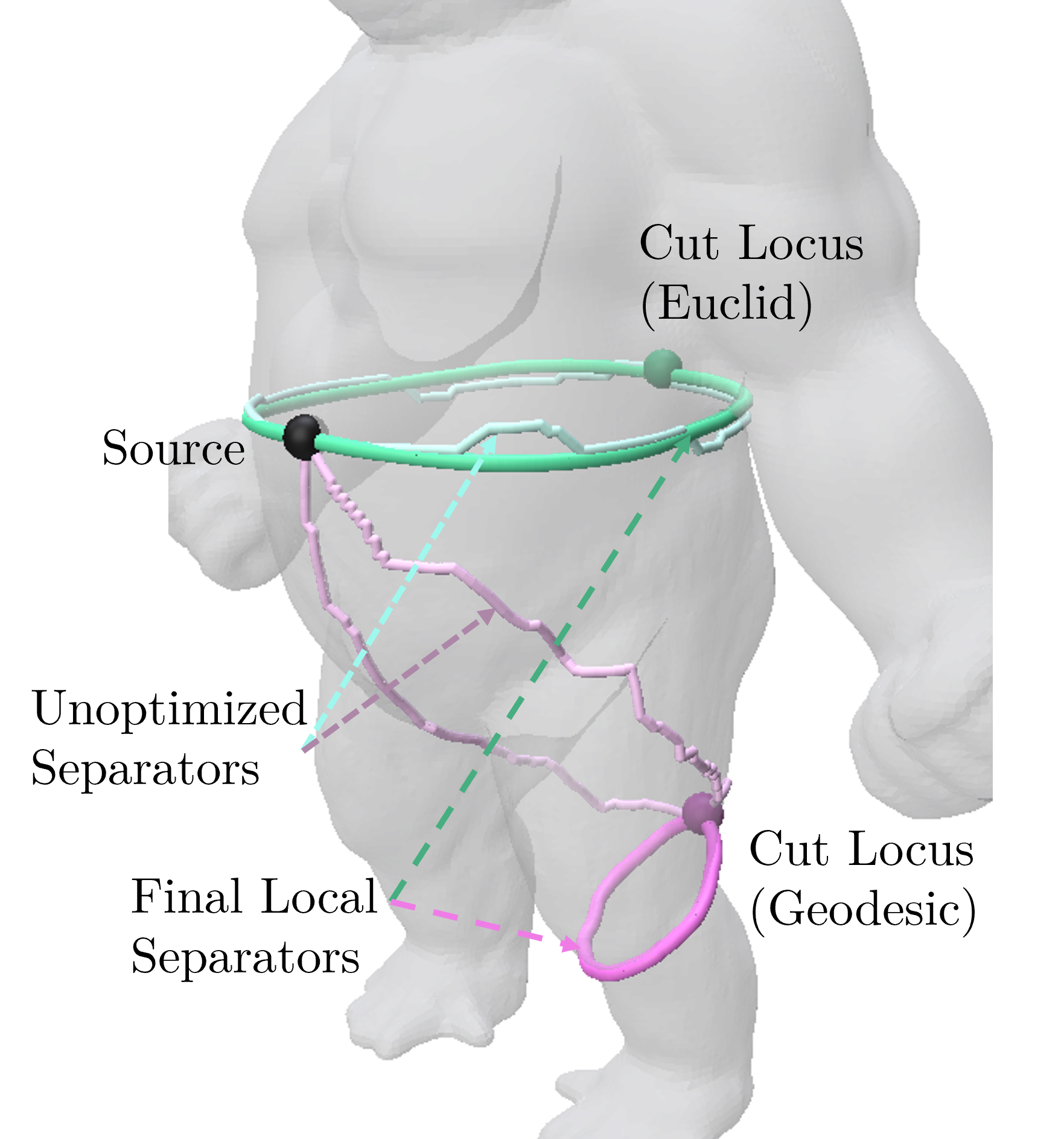}
    \label{fig:target-cl-chose}
\end{wrapfigure} 
Following \cite{origLS}, we select the target cut locus $t$ as the one with the smallest Euclidean distance to the source $s$, ensuring that the cut locus lies near a significant feature. Since intrinsic triangulation schemes lack explicit vertex positions in $\mathbb{R}^3$, we also test selecting $t$ using the smallest geodesic distance (see App.~\ref{sec:geodesic-vs-euclid}).

\subsubsection{Approximate Path Construction}
\label{sec:trace-path}
For \cscd-M, we search the neighborhood of the target cut locus for two incoming directions, separated by the cut loci graph (inset). 

For \cscd-PC, a similar strategy is followed but with the additional requirement that the gradient directions are opposite. 
\begin{wrapfigure}[10]{r}{0pt}
    \includegraphics[width=0.38\linewidth]{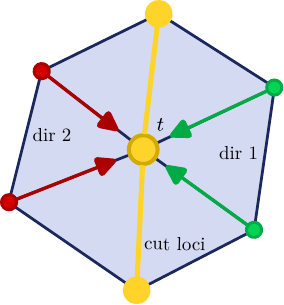}
    \label{fig:dir-approach}    
\end{wrapfigure}
Two paths are constructed by greedily following vertices (or points) with the minimum geodesic distance from the source $s$, ensuring convergence at $s$. 
In cases where the paths meet at an intermediate vertex $v$, they are truncated at $v$. The concatenation of these paths forms a loop; 
Fig.~\ref{fig:curve-shorten} shows the approximate paths in red and lime green. The cut loci is visualized in yellow.

\subsubsection{Optimizing the Loop}
\label{sec:loop-optim}
At this stage, the approximate loop only satisfies property \textbf{P1}. We aim to shorten the loop around the feature (see the cyan loop in Fig.~\ref{fig:curve-shorten}). For \cscd-M, this is achieved using an edge flip procedure~\cite{sharp} in the intrinsic triangulation, while for \cscd-PC, an optimization-based framework from~\cite{yuan_opt} is employed (see App.~\ref{sec:append-edge-flip-optim} and App.~\ref{sec:append-curve-shorten-optim}).

\subsubsection{Constraining the Loop}
\label{sec:bounding-sphere-constraint}

\begin{wrapfigure}[17]{r}{0pt}
    \includegraphics[width=0.5\linewidth]{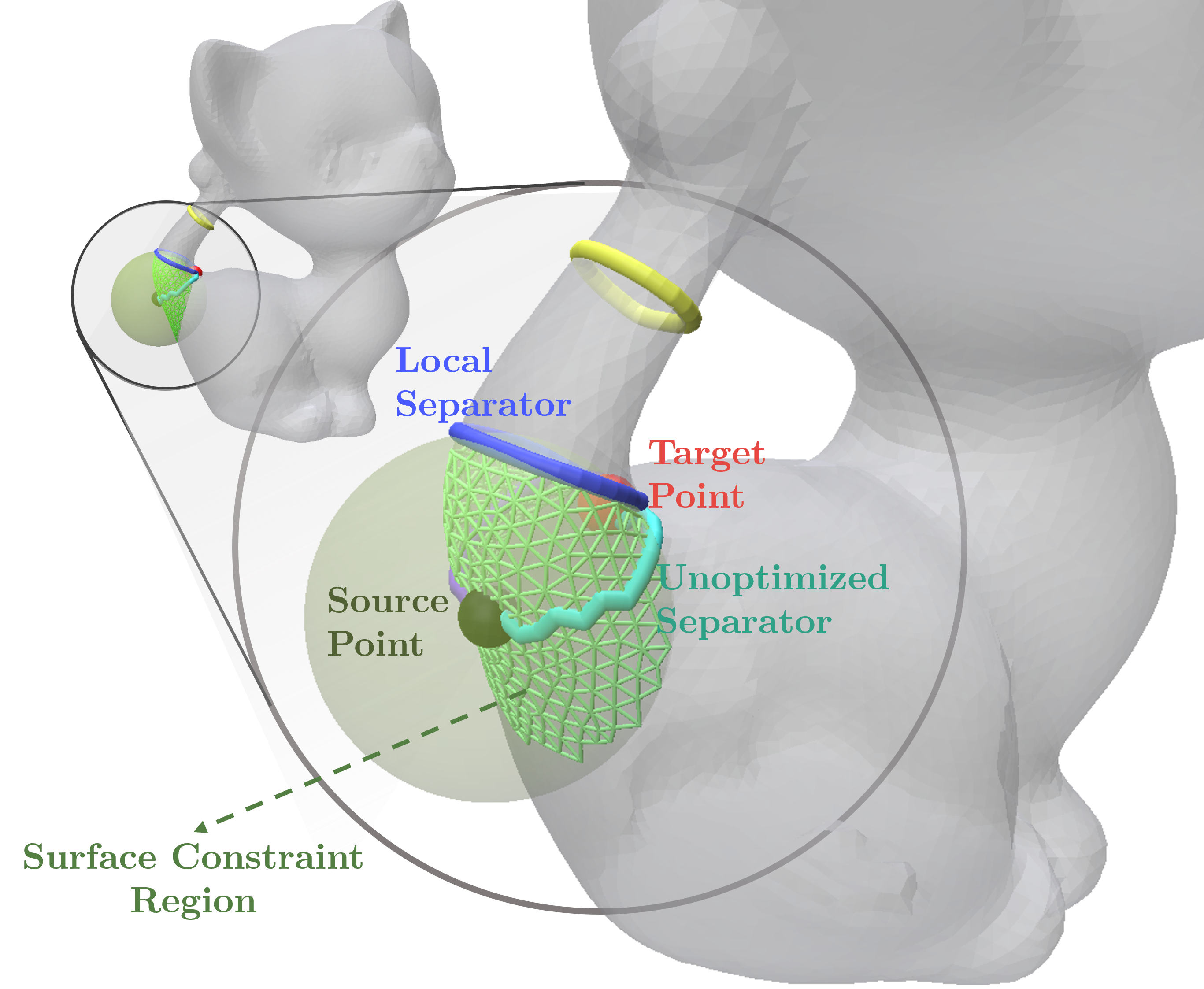}
    \caption{Constraint region for the loop optimization procedure. The green sphere shows the Euclidean sphere constraint.}
    \label{fig:sphereC}
\end{wrapfigure}
Simply shortening the curve yields a local geodesic loop that can drift significantly from the initial path. For example, a loop drawn at the bottom of a cone may slide upward toward the tip, which is undesirable for curve skeletonization.

Two observations guide our constraint: (1) the target cut locus is selected based on Euclidean proximity, and (2) the final separator should not lie farther from the source than the most distant point on the approximate path. Thus, we apply a bounding sphere constraint centered at the source $s$ with a radius equal to the Euclidean distance from $s$ to the furthest point on the approx. separator. In Fig.~\ref{fig:sphereC}, with the constraint, the optimized separator (dark blue) remains in place, whereas without it the separator shifts upward (yellow).

In \cscd-M, the constraint restricts edge flips for vertices outside the sphere. In \cscd-PC, it is imposed as an interior point constraint in the optimization energy:
\begin{equation} \label{eq:optim_loss_main_text}
    \mathcal{L} = \sum_{i = 1}^n H\left(\| x_i - x_{i+1}\|_2\right)
     + \sum_{i=1}^n \lambda_i \max\left(0, \left\|x_i - x_s\right\|_2 - r \right)^2,
\end{equation}
where $x_s$, $r$, and $H$ denote the source point, the Euclidean radius, and a kernel function, respectively.

\subsubsection{Sampling the Separators}
\label{sec:adaptive-sampling}
We avoid sampling regions that produce similar local separators by employing an adaptive sampling technique based on geodesic distances. The distance from each vertex to the constructed separators and sampled points is computed. Regions with larger distances are more likely to yield unique separators. The probability for a vertex $i$ is given by $p_i \propto \exp\left(d_i\right) - 1$, where $d_i$ is the minimum geodesic distance from vertex $i$ to the existing separators and source points. The exponential weighting tends to emphasize separators around distant, sharp boundary features.

\subsection{Stage 2: Constructing the Curve Skeleton}

\subsubsection{Scoring the Separators}
\label{sec:sep-score}
We define a score for each separator to decide among overlapping candidates. Inspired by LS, a good local separator balances the two halves of the surface constraint region mentioned in sec.~\ref{sec:bounding-sphere-constraint}. Instead of simply counting nodes, we weigh based on the ratio of the surface areas of the two components and penalize longer loops. The final score is $s_i = \frac{A_{1,i}}{A_{2,i} \cdot l_i},$
where $A_{1,i}$ and $A_{2,i}$ (with $A_{2,i} > A_{1,i}$) are the surface areas of the two components, and $l_i$ is the loop length.

\subsubsection{Pruning Bad Separators}
\label{sec:sep-prune}
Tiny separators that do not enclose a significant feature are pruned using a length threshold $\tau = 3 \times \bar{d_{ij}}$, where $\bar{d_{ij}}$ is the average edge length. Separators forming handles rather than properly encircling features are removed by discarding those whose centroids lie outside the 3D shape.

\subsubsection{Packing the Separators}
\label{sec:set-packing}
After scoring, set packing is performed to suppress overlapping separators. We normalize the weights based on the opportunity cost of retaining one separator over another and greedily select those with the highest normalized weights. Separators that do not overlap are retained automatically.

For \cscd-M, overlapping separators are identified by testing intersections of piecewise linear curves within each face (\textbf{See Supplementary Sec.~\ref{sec:append-derivation-intersect} for Derivation on determining intersection within a face }). For \cscd-PC, a distance threshold between points is used. With the selected set of non-overlapping separators, we proceed to construct the curve skeleton graph.

\subsubsection{Assigning Regions}
Neighboring regions of the object are assigned to each local separator by computing the geodesic distance from every vertex (or point) to each separator, followed by a Voronoi partitioning.

\begin{figure}[htbp]
    \centering
    \begin{subfigure}[b]{0.3\linewidth}
    \includegraphics[width=\linewidth]{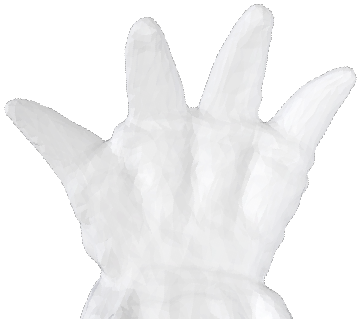}
        \caption{}
    \end{subfigure}
    \hfill
    \begin{subfigure}[b]{0.3\linewidth}        \includegraphics[width=\linewidth]{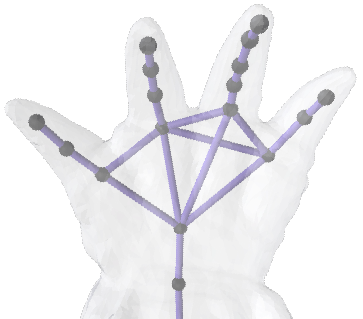}
        \caption{}
    \end{subfigure}
    \hfill
    \begin{subfigure}[b]{0.3\linewidth}
\includegraphics[width=\linewidth]{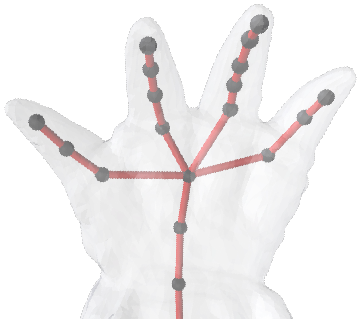}
        \caption{}
    \end{subfigure}
    \caption{Cliques in the \texttt{armadillo} hand and their removal. (a) Original hand; (b) With cliques; (c) Cliques replaced by central star-like nodes.}
    \vspace{-1em}
    \label{fig:cliques}
\end{figure}

\subsubsection{Constructing the Graph}
The centroids of the regions form the nodes of the curve skeleton graph. Nodes corresponding to neighboring regions are connected. When a region neighbors more than two others, resulting cliques are simplified by converting them to a star formation, using the centroid as the central node and removing redundant edges (see Fig.~\ref{fig:cliques}).

\subsection{On the Discrete Nature of the Realizations}
\label{sec:disc-nature-realization}

Manifold representations are inherently discrete in computers; thus, our realizations of \cscd are discrete and involve tradeoffs similar to those in adapting geodesic paths (e.g., straightest vs. shortest paths). While LS relies solely on node and edge data, our framework benefits from additional face-level information for meshes, allowing for interpolation across faces. In point clouds, local separators are constructed using dynamically computed neighborhood information. Although the complete graph is not built initially, many computations are reused, and techniques such as MLS or tangent space smoothing can refine the cut loci and loop approximations. For other discrete representations (e.g., digital surfaces), our method remains discrete, though it could incorporate continuous-level corrections if available.

%% file: sections/4_results.tex
\section{Results}

\subsection{Results of the Improved Cut Loci Identification}
\label{sec:results-cl}

\begin{figure}[ht]
    \centering
    \begin{subfigure}[b]{0.49\linewidth}
        \centering
        \includegraphics[height=2.5cm]{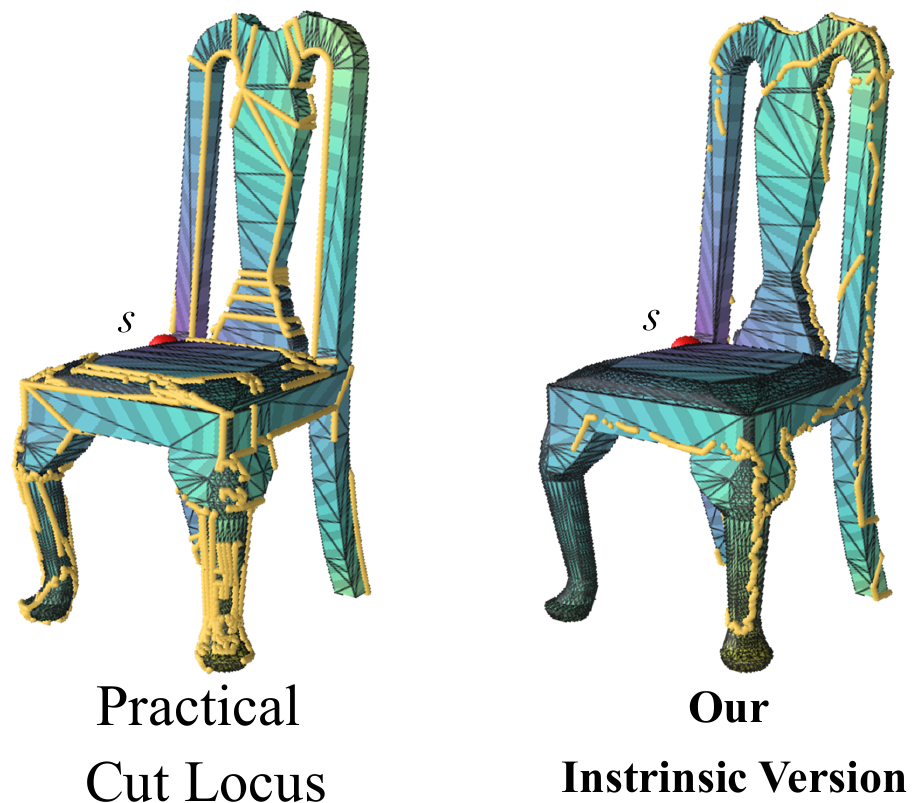}
        \caption{Mesh}
        \label{fig:res-comparion-cut-locus-identify}
    \end{subfigure}
    \hfill
    \begin{subfigure}[b]{0.49\linewidth}
        \centering
        \includegraphics[height=2.5cm]{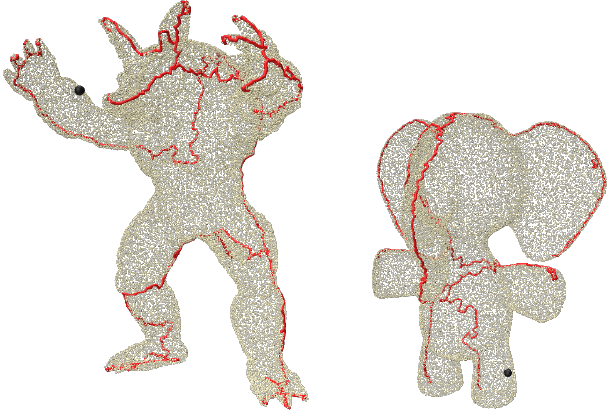}
        \caption{Point Cloud}
        \label{fig:res-cl-pc}
    \end{subfigure}
    \caption{(a) Comparison of our (right) estimated cut loci (\textit{in yellow}) versus the previous approach (left)~\cite{cutLocusComp} (of source $s$) on \texttt{chair} from \texttt{Thingi10k}. \textbf{Note:} Our adaptation generates robust output by selectively identifying points on the true cut loci, thereby significantly minimizing false negatives; (b) our novel cut loci identification algorithm applied to point clouds. Source point is shown in black and the resulting cut loci are rendered as red curves.}
    \label{fig:combined-cut-locus}
\end{figure}

\begin{figure*}
    \centering
    \includegraphics[height=6cm]{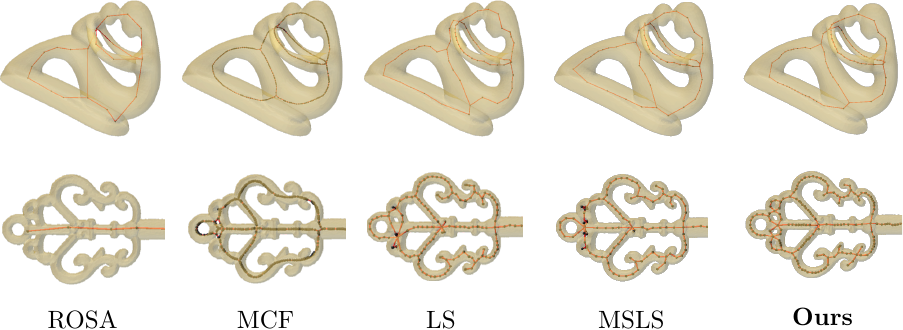}
    \caption{\textbf{Qualitative results of our method:} ROSA~\cite{rosa} struggles to capture mesh details, while MCF~\cite{tag_mc} produces overly smooth skeletons. \cscd-M, LS~\cite{origLS} and MSLS~\cite{fastLS} yield comparable results on meshes; however, our method correctly captures details, as seen on the copper key. In \texttt{fertility}, our approach results in smoother skeletons compared to LS and MSLS.}
    \label{fig:qual-res-all}
\end{figure*}

We show the results of our cut locus identification algorithm in Fig.~\ref{fig:res-comparion-cut-locus-identify} and Fig.~\ref{fig:res-cl-pc}. Our adaptation produces robust output that selectively chooses the points on the cut loci, thereby significantly reducing the false negatives. For details see Appendix~\ref{sec:practical-cl-adapt}.

\subsection{Curve Skeletonization on Meshes}
\label{sec:results-mesh}

\emph{\textbf{General performance of \cscd-M}:} We evaluate and compare our method to \cite{origLS, fastLS, rosa, tag_mc} on diverse objects mostly from the Stanford 3D Library, Artec 3D Scans, and Thingi10k datasets. Figs.~\ref{fig:main_figure} and \ref{fig:qual-res-all} illustrate that our curve skeletonization is topologically correct and faithfully follows the object geometry. In comparison to ROSA and MCF—which often miss key features—our method (and LS) retains more details. Notably, our skeletons are smoother and yield better-centered nodes without additional smoothing; we suspect this is due to weighing centroid computations with vertex and face areas. In the case of \texttt{Copper Key}, our method uniquely captures the intricate design (See Fig. \ref{fig:main_figure} (b)).

\begin{figure}[hbt]
    \centering
    \includegraphics[width=\linewidth]{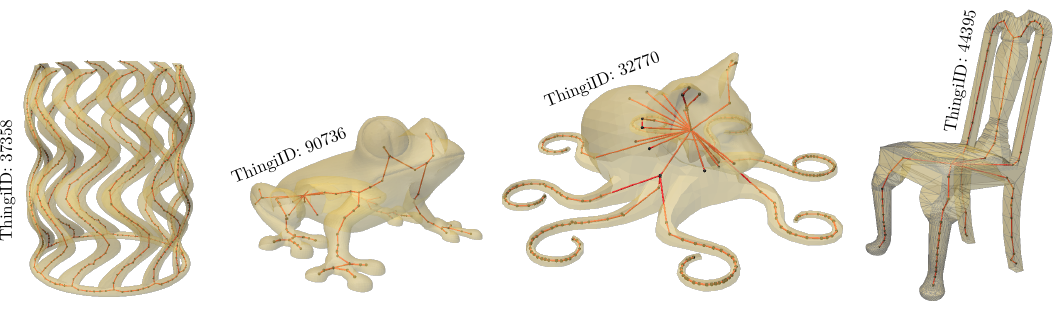}
    \caption{\textbf{Qualitative results of \cscd-M on Thingi10k.} \cscd-M performs well even on poorly triangulated meshes.}
    \label{fig:res-thingi10k-qual}
\end{figure}

\vspace{1em}
\noindent\emph{\textbf{Performance on poorly triangulated meshes}:} Our intrinsic triangulation scheme, based on the Integer coordinate system~\cite{gillespie2021integer}, uses intrinsic edge flips to enforce Delaunay conditions. As shown in Fig.~\ref{fig:res-thingi10k-qual}, our method reconstructs skeletons effectively on such meshes; Tab.~\ref{tab:quant-result-all}. Also see Fig.~\ref{fig:res-noisy-mesh} for results on noisy meshes. Fig.~\ref{fig:res-mesh-resolution}.

\begin{figure}[htb]
    \centering
    \includegraphics[width=\linewidth]{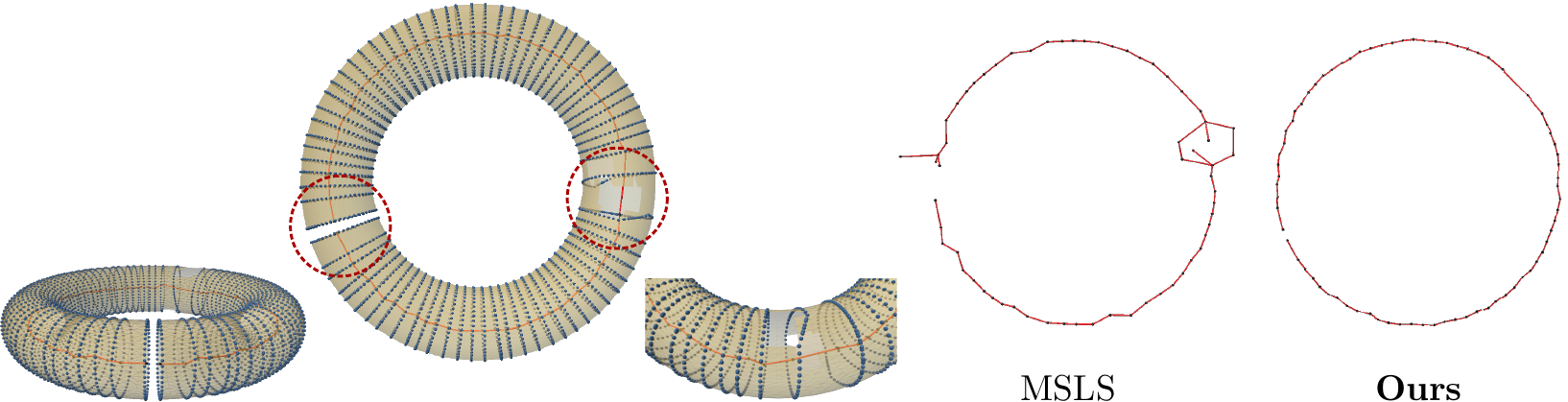}
    \caption{\cscd-M and MSLS on a \textbf{torus with holes}. Left panel, \textbf{\cscd-M local separators on the torus} -- highlighting how the separators go around the holes. Right panel, \textbf{comparison of the curve skeleton} obtained by MSLS and Ours (\cscd-M).}
    \label{fig:res-holes-result-sec}
\end{figure}

\begin{table}[hb!]
    \centering
    \caption{\emph{(Truncated)} Convolutional Surfaces reconstruction error ($\times 10^{-3}$) for objects using skeletons from \cscd-M (Ours), MCF~\cite{tag_mc}, LS~\cite{origLS} and MSLS~\cite{fastLS}.  Average is computed over the subset here. Complete Table~\ref{tab:quant-result-all}.}
    \resizebox{0.38\textwidth}{!}{
    \begin{tabular}{l|c|c|c|c}
    \toprule
       Object  &  LS & MSLS & MCF & \cscd-M \textbf{(Ours)}\\
    \midrule   
        \texttt{Copper-key} & $06.13 $ & $\mathbf{04.60}$ & $6.02$ & \underline{$04.70$} \\
        \texttt{rocker-arm}  & $26.30 $ & $\mathbf{24.40}$ & $25.30$ & \underline{$24.90$}\\
        \texttt{neptune}  & \underline{$04.16$} & $04.82$  & $10.62$ & $\mathbf{03.80}$\\
        \texttt{TID:44395}  & $\mathbf{09.56}$ & $10.70$ & $16.80$  & \underline{$10.07$}\\
        \texttt{TID:40987} & \underline{$09.29$} & $09.81$ & $11.36$ & $\mathbf{08.32}$ \\
        \texttt{TID:133568} & $\mathbf{04.75}$ & $05.49$ & $05.51$ & \underline{$04.99$} \\
    \midrule
        \textbf{Average} & $10.03$ & $9.97$ & $12.60$ & $\mathbf{9.46}$ \\
        
    \bottomrule
    \end{tabular}
    }
    \label{tab:quant-result}
\end{table}

\noindent\emph{\textbf{Performance on meshes with holes}:} We evaluate on a torus mesh with three holes—two partial and one fully disconnecting the shape—as a controlled test case. Our method is the only one to perform reliably (Fig.~\ref{fig:res-holes-result-sec}); LS fails to produce output; MSLS recovers incorrect topology.

\vspace{1em}
\noindent\emph{\textbf{Quantitative performance of \cscd-M}:} 
Due to the lack of a standard quantitative metric for curve skeletonization, we propose a reconstruction loss based on convolutional surfaces~\cite{suarez2019modeling}, where the error at vertex \( i \) is \( \epsilon_i = \min_{\hat{p} \in \hat{\mathcal{M}}} \|\mathbf{p}_i - \hat{p}\|^2 \), with \( \hat{p} \) any point on the reconstructed mesh \( \hat{\mathcal{M}} \) (see App.~\ref{sec:append-experiment-setup}). As shown in Table~\ref{tab:quant-result}, our method outperforms LS, likely due to smoother, more centered skeletons.

\vspace{1em}
\noindent\emph{\textbf{Inference times}:} Our method on average performs faster than LS on our subset of objects ranging in different complexities (see Table~\ref{tab:timing-inference-res-sec}). For additional details see App.~\ref{sec:append-inference-times}.

\subsection{Curve Skeletonization on Point Clouds}
\begin{figure}[htp]
    \centering
    \includegraphics[width=0.9\linewidth]{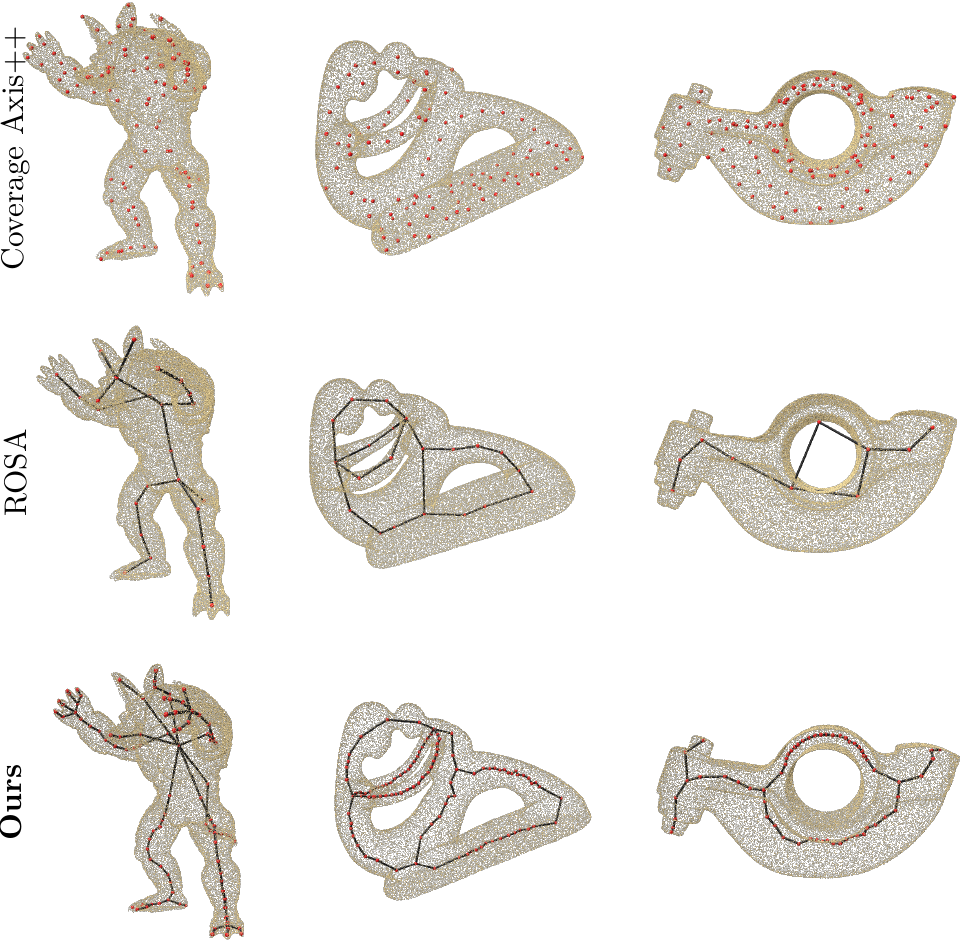}
    \caption{\textbf{\cscd-PC vs ROSA~\cite{rosa} vs CA++ [Eurographics'24]\cite{wang2024coverage}:} CA++ fails to generate a valid skeleton (since it's a MAT inspired algorithm). Our method captures object details better, yielding more nodes and centered skeletons compared to ROSA.}
    \label{fig:cscd_pc_vs_rosa_res}
\end{figure}

\begin{figure}[htp]
    \centering
    \includegraphics[width=\linewidth]{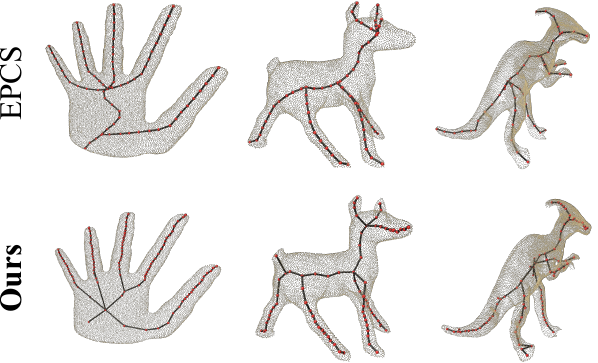}
    \caption{\textbf{\cscd-PC vs EPCS [CAG'23]:} Our method captures object details better,  compared to EPCS~\cite{LI2023209}. In \texttt{hand} (left), our skeleton is centered in the palm and shows skeleton consistent with square shape. In \texttt{deer} (middle), we capture the snout and the tail. In \texttt{dino} (right), EPCS fails to capture the curvature of the arm completely while our skeleton follows the arm.}
    \label{fig:cscd_pc_vs_epcs_res}
\end{figure}

We demonstrate a proof-of-concept realization of \cscd on point clouds (\cscd-PC), which captures detailed, well-centered skeletons (Figs.\ref{fig:cscd_pc_vs_rosa_res},\ref{fig:cscd_pc_vs_epcs_res}), outperforming ROSA, CA++, and EPCS, which yield coarser skeletons with fewer nodes.

\subsection{Downstream Applications}

\subsubsection{Object Classification}
We compare curve skeletonization methods for object classification on the Princeton Shape Benchmark~\cite{shilane2004princeton}. A global shape embedding is constructed via a histogram of the Shape Diameter Function (SDF)~\cite{shapira2008consistent}, and the 1D Wasserstein distance between histograms is used for comparison, ensuring robustness against discretization. Our method outperforms contemporary approaches (see, Table~\ref{tab:res-shape-classification}).

\subsubsection{Object Segmentation}
We also evaluate our outputs for unsupervised object segmentation using SDF~\cite{shapira2008consistent}, yielding consistent results robust to pose variations (Fig.~\ref{fig:shape-segmentation-append}).

\noindent \textbf{Note:} App. for more results, ablations, downstream task.

\begin{table}[t]
    \centering
    \caption{\textit{(Truncated)} Runtime analysis (in secs.) for \cscd-M (Ours), LS~\cite{origLS} (TOG'21) and MSLS~\cite{fastLS}. Results are based on $N = 3K$ (with three cases using $N=4K$ due to mesh complexity). LS times for \texttt{gorilla} are omitted from averaging due to excessive computation time. Averaging is computed over the complete table.  Complete Tab.~\ref{tab:timing-inference}}
    \resizebox{0.47\textwidth}{!}{

    \begin{tabular}{l|c|c|c|c|c}
    \toprule
    Object           & $|V|$   & $|F|$    & LS               & MSLS     & \cscd-M (Ours)     \\
    \midrule
    \texttt{TID:44395}   & 2948    & 5900     & 6.78             & 2.14     & 37.65              \\
    \texttt{fertility}   & 4494    & 9000     & 9.39             & 2.79     & 41.43              \\
    \texttt{TID:32770}   & 20125   & 40246    & 1889.70          & 13.20    & 225.17$^*$         \\
    \texttt{gorilla}     & 48762   & 97520    & $\geq 2400^\dagger$ & 32.46    & 532.16             \\
    \texttt{armadillo}   & 49990   & 99976    & 994.75           & 33.70    & 495.33             \\
    \texttt{garuda-vishnu}& 49972  & 100084   & 292.43           & 30.13    & 609.30$^*$         \\
    \texttt{neptune}     & 50000   & 100008   & 886.68           & 31.53    & 611.66$^*$         \\
    \midrule
    \textbf{Average}     &         &          & 471.17           & 16.14    & 277.99             \\
    \bottomrule
    \end{tabular}
    
    }
    \label{tab:timing-inference-res-sec}
\end{table}

\begin{table}[t!]
    \centering
    \caption{\textbf{Application I: Shape Classification:} For subset of classes from the Princeton Shape Benchmark dataset~\cite{shilane2004princeton}.}
    \resizebox{0.26\textwidth}{!}{
    \begin{tabular}{c|c|c|c}
        \toprule
        Metric & LS & MSLS & \cscd-M \\
        \midrule
        Accuracy & $0.63$ & $0.74$ & $\mathbf{0.79}$ \\ 
        F$1$ Score & $0.59$ & $0.76$ & $\mathbf{0.80}$ \\
        \bottomrule
    \end{tabular}
    }
    \label{tab:res-shape-classification}
\end{table}

%% file: sections/6_conclusions.tex
\section{Conclusion and Future Work}

We introduced \cscd, a general framework for curve skeletonization on continuous manifolds that generalizes LS. Its effectiveness is demonstrated through implementations on meshes (\cscd-\texttt{M}) and point clouds (\cscd-\texttt{PC}). \cscd-\texttt{M} is the first intrinsic curve skeletonization method and shows robust performance across diverse meshes, with results that are comparable or superior to the state-of-the-art LS. Meanwhile, \cscd-\texttt{PC} provides a compelling proof-of-concept for the framework's generalizability. Our realizations are intended as starting points, with results on meshes and point clouds. Future work could improve individual modules for improved speed, robustness, and performance, or extend the framework to other representations. We plan to release the source code post acceptance. 

\section{Acknowledgments}

We would like to thank Rahul Narain for the insightful initial discussions on shape representation, which laid the foundation for this work. We would also like to thank Cyrin Neeraj for his help on results presentation.

%% file: sections/supplementary.tex
\clearpage
\renewcommand{\thesection}{\Alph{section}}
\setcounter{page}{1}
\appendix

\onecolumn{
{
        \centering
        \Large
        \textbf{\title}\\
        \vspace{0.5em}Supplementary Material \\ Generalizing Curve Skeletonization to Continuous Domains\vspace{1pt}\\
}

\vspace{3em}
\startcontents[sections]
\printcontents[sections]{l}{1}{\setcounter{tocdepth}{2}}

\clearpage

\section{Implementation Details}
The code for meshes is implemented in C++, using the \texttt{IGL} \cite{libigl} and \texttt{geometry-central} \cite{geometrycentral} package for geometry processing. 
The code for point clouds is implemented in Python, using \texttt{PyIGL} \cite{libigl}, \texttt{Open3D} \cite{open3d} and \texttt{PotPourri3D} \cite{geometrycentral} for geometry processing. KDTree from \texttt{scikit-learn} \cite{scikit-learn} is used to implement MLS.

\section{Pseduocode}
\label{sec:pseudocode}

We provide in the following section, the pseudocode for our algorithm. The entire framework can be divided into 2 broad stages as depicted in \cref{fig:proc-diag}. Stage 1 is to find a set of local separators on the 3D object. Given an input 3D object, we sample a point on the surface. We then calculate the geodesic distance from the source to all other points and identify the target cut locus for the source. We then trace an approximate path, which is then optimized using curve shortening. In Stage 2, we first score the separators, and then select an optimal set of separators from overlapping separators. These steps are followed by assigning neighbouring regions to local separators, calculating the centroid of each region and connecting centroids of neighbouring regions to form the final curve skeleton.
\begin{algorithm}[H]
  \caption{Obtain curve skeleton from local separator}
  \label{alg:get-curve-skel}
  \begin{algorithmic}[1]
    \Require A 3D mesh, the number of local separators to construct $N$
    \Ensure A curve skeleton $\mathcal{C}$ (graph with interior‐mesh nodes)
    \State LS $\gets \{\}$
    \State $P_V \gets \mathbb{U}_V$ \Comment{Uniform sampling over all mesh vertices}
    \For{$t \gets 0,1,\dots,N-1$}
      \State $s \sim P_V$ \Comment{Sample a source vertex}
      \State $\mathrm{LS}_t \gets \texttt{local\_separator}(s)$ \Comment{Alg.~\ref{alg:get-ls}}
      \State Update $P_V$ via Sec.~\ref{sec:adaptive-sampling}
      \State Compute weight $w_t$ via Sec.~\ref{sec:sep-score}
      \State LS $\gets$ LS $\cup \{(\mathrm{LS}_t, w_t)\}$
    \EndFor
    \State Prune bad local separators (Sec.~\ref{sec:sep-prune})
    \State LS $\gets \texttt{set\_packing}(LS)$ \Comment{Sec.~\ref{sec:set-packing}}
    \State Assign nearest separator to each mesh vertex
    \State Compute centroids for each region
    \State Connect neighboring regions and remove cliques to form $\mathcal{C}$
  \end{algorithmic}
\end{algorithm}

\begin{algorithm}[H]
  \caption{Obtain a local separator}
  \label{alg:get-ls}
  \begin{algorithmic}[1]
    \Require A mesh with a source point $s$
    \Ensure A set of points in $\mathbb{R}^3$ as the local separator
    \State Compute distance field $D$ (e.g.\ via the Heat Method \cite{Crane:2017:HMD})
    \State Identify cut loci using the algorithm of \cite{cutLocusComp} (see Sec.~\ref{sec:cut-locus-identify})
    \State Select target cut locus $t$ by the locality constraint (min‐Euclid or min‐geodesic; Sec.~\ref{sec:target-cl})
    \State Determine incoming directions $v_1, v_2$ (Sec.~\ref{sec:trace-path})
    \State Trace two paths from $t$ to $s$ using Alg.~\ref{alg:trace-dir}, yielding \texttt{Path1}, \texttt{Path2}
    \State Connect \texttt{Path1} and \texttt{Path2} into a closed loop \texttt{loop}
    \State Optimize \texttt{loop} using intrinsic edge‐flips (\cite{sharp}) or the procedure in \cite{yuan_opt}
  \end{algorithmic}
\end{algorithm}

\begin{algorithm}[H]
  \caption{Trace direction to source $s$}
  \label{alg:trace-dir}
  \begin{algorithmic}[1]
    \Require A mesh with distance field $D$, a target cut locus $t$, source $s$, and an incoming direction $v_i$
    \Ensure A vertex path from $v_i$ to $s$
    \State $\mathit{Path} \gets [\,v_i\,]$
    \While{$\mathrm{last}(\mathit{Path}) \neq s$}
      \State $\mathit{minD} \gets +\infty$
      \State $\mathit{bestV} \gets \mathrm{null}$
      \ForAll{$n \in \mathrm{adj}(\mathrm{last}(\mathit{Path}))$}
        \If{$D[n] < \mathit{minD}$}
          \State $\mathit{minD} \gets D[n]$
          \State $\mathit{bestV} \gets n$
        \EndIf
      \EndFor
      \State Append $\mathit{bestV}$ to $\mathit{Path}$
    \EndWhile
  \end{algorithmic}
\end{algorithm}

\section{Derivation for determining intersection within a face of a mesh}
\label{sec:append-derivation-intersect}
We derive below the equation that needs to be evaluated to check whether two barycentric vectors are intersecting within the face of a mesh. This is used to check for the overlap of local separators in \cscd-M. 

Let the the first vector be given by two barycentric points $(u_1, v_1, w_1)$, $(u_2, v_2, w_2)$, and the second vector by given by another set of two barycentric points $(u_3, v_3, w_3)$, $(u_4, v_4, w_4)$.

Now, if we parameterize the lines using $t_1$ and $t_2$, we have:
\begin{align}
    L_1(t_1) &= \left(u_1 + t_1 (u_2 - u_1), v_1 + t_1(v_2 - v_1), w_1 + t_1(w_2 - w_1)\right) \\
    L_2(t_2) &= \left(u_3 + t_2 (u_4 - u_3), v_3 + t_2(v_4 - v_3), w_3 + t_2(w_4 - w_3)\right)
\end{align}

At the intersection, $L_1(t_1) = L_2(t_2)$.
For barycentric coordinates, we have $u + v + w = 1$ for all $u, w, v \in \mathbb{R}$. Therefore, we can pick $u, v$ and set $w = 1 - u - v$. Now, the above relation boils down to two linear equations with two unknowns $t_1$ and $t_2$:
\begin{align}
    u_1 + t_1(u_2 - u_1) &= u_3 + t_2(u_4 - u_3) \\
    v_1 + t_1(v_2 - v_1) &= v_3 + t_2(v_4 - v_3) \\ 
\end{align},
which has exactly one solution, 
\begin{align}
    t_1 &= \frac{u_1 (v_4 - v_3) + u_3 (v_1 - v_4) + u_4 (v_3 - v_1)}{(u_1 - u_2) (v_4 - v_3) - (u_4 - u_3) (v_1 - v_2)} \\
    t_2 &= \frac{u_1 (v_2 - v_3) + u_2 (v_3 - v_1) + u_3 (v_1 - v_2)}{(u_1 - u_2) (v_4 - v_3) - (u_4 - u_3) (v_1 - v_2)}
\end{align},

where the denominator is the same and is zero if and only if the two lines are parallel.

If $0 \leq t_1 \leq 1$, then the intersection is on the line segment between the specified points of the first line. Similarly for $t_2$. 

The point lies within the face if $0 \leq u \leq 1, 0 \leq v \leq 1$ and $0 \leq w \leq 1 \iff 0 \leq u + v \leq 1$, which is always the case here since the point lies at the edge of the face.

\section{Evaluation metric}
\label{sec:append-experiment-setup}

Due to the unclear definition of curve skeletons for 3D objects, there is no single agreed-upon evaluation metric for quantifying the quality of the skeleton. Here, we choose to evaluate the reconstruction quality of the mesh. Each node of the skeleton corresponds to a sphere of a radius $r$, which we use to perform convolutional surfaces~\cite{suarez2019modeling} to reconstruct the original mesh. This metric is well-suited since it requires that the nodes not be centered (due to the isotropic nature of the radius), but also capture the finer details of the mesh.    

We describe briefly the evaluation process given the reconstructed mesh obtained from applying convolutional surfaces:
We compute the Euclidean distance between points on the real object against a location on the reconstructed object. This differs from the standard Chamfer distance, which typically compares two point clouds. In this case,  the evaluation compares a set of points -- the vertices of the original mesh -- against a mesh representing the reconstructed object. Notably, the closest point on the reconstructed mesh may not always be a vertex; it could instead lie on a face. 
 
\begin{equation}
    \epsilon_i = \min_{\hat{p} \in \hat{\mathcal{M}}} \|\mathbf{p_i} - \mathbf{p}\|^2
\end{equation}
where $\epsilon_i$ is the error for vertex $i$, $\mathbf{p_i}$ is the position of vertex $i$, and $\mathbf{\hat{p}}$ is a point on the reconstructed mesh $\hat{\mathcal{M}}$.

\section{Curve Shortening using Edge Flip framework}
\label{sec:append-edge-flip-optim}

\subsection{Optimizing the loop}
For \cscd-M, we use the curve shortening framework described in~\cite{sharp} as it works out-of-the-box for intrinsic triangulation schemes and is fairly robust. In intrinsic triangulation schemes, the diagonal edges of a quadrilateral, formed by two adjacent triangles, may be flipped. The geodesic curve shortening algorithm utilizes this detail and works by iteratively flipping edges in a wedge, until the interior angle of the two edges connecting two non-adjacent vertices on the path is greater than $\pi$. Therefore, one can begin with an approximate path given by a series of vertices and iteratively smoothened it to the final geodesic path by flipping edges connecting the vertices. In the case of loops, there is no fixed start and end vertices, which causes the loop to become geodesic loops.

\subsection{Constraining the Loop}

To prevent the optimized loop from moving too far from the initial approximate loop, we apply a locality constraint. This is done through restricting the edge flipping procedure to operate only upon certain vertices that are within the locality constraint. This would ensure that the curve shortens and smoothens within the locality but doesn't slide away.

\section{Curve Shortening using optimization based framework}
\label{sec:append-curve-shorten-optim}
\subsection{Optimizing the Loop}

We employ the method described in ~\cite{curveShortening}. The method proposes it as an optimization problem and solves it using projected gradient descent. We discuss below the method:

Let us take a loop $\mathbb{L}$ with $n$ points $p_1, p_2, p_3, \dots, p_n$ such that there is connectivity between consecutive points and between $p_n, p_1$. Assuming proper initialization, it has be shown that the loop can be converted into a geodesic loop by (i.e, geodesic minimization is equivalent to) minimizing the following loss~\cite{curveShortening}:
\begin{equation} \label{eq:optim_loss}
    \mathcal{L} = \sum_{i = 1}^n H\left(\| x_i - x_{i + 1}\|_2\right)
\end{equation}

where $x_i$ is the Euclidean position of point $p_i$, and $H$ is a convex kernel function. The kernel function $H(s) = e^{s^2} - 1$ has been shown to work very effectively, leading to quick convergence rates \cite{curveShortening}, so we use the same kernel. 


In order to optimize the loop $\mathbb{L}$, the gradient of $\mathcal{L}$ is projected onto the surface of the object as:
\[
   \bar{\nabla}_{i}\mathcal{L} =  \nabla_{i}\mathcal{L} - \nabla_{i}\mathcal{L} \cdot N_i
\]
where $N_i$ is the normal to point $p_i$. After each step of the optimization, the points are reprojected onto the surface of the point cloud using MLS \cite{Fleishman2005RobustML}. The optimization runs until we reach a sufficiently small step size or a sufficiently small gradient norm.

\subsection{Constraining the Loop}

We implement the Euclidean distance constraint as a penalty to the original objective~\eqref{eq:optim_loss}:
\begin{equation}
    \tilde{\mathcal{L}} = \mathcal{L} + \sum_{i=1}^n \lambda_i \max\left(0, \left|\left|x_i - x_s\right|\right|_2 - r \right)^2
\end{equation}
where $r = \left|\left|x_p - x_s\right|\right|_2$, i.e., the Euclidean distance between the source point and the most distant point, and $\lambda_i$ are hyperparameters for weighing the penalty function. One could apply interior point methods to simultaneously fit $\lambda_i$ and the positions $x_i$, but we find that simply setting $\lambda_i = 10$ works pretty well.
This constraint prevents the loop (separator) from moving too far from the initial region or collapsing onto a single point.

For \cscd-M, we also experimented with the optimization based framework, but could not get it work as efficiently as mentioned in~\cite{yuan_opt}. The optimization based framework requires the position in $\mathbb{R}^3$, but provides greater flexibility than the edge flip method since does not depend upon the triangulation scheme.

\section{Our adaptation of the Practical Cut Locus Identification Method}
\label{sec:practical-cl-adapt}
\subsection{Our adaptation on intrinsic meshes}

We adapt the practical cut locus mentioned in Ref.~\cite{cutLocusComp} for intrinsic mesh triangulation. The key ingredients in the algorithm are the computation of the geodesic distance, the computation of the gradients of the distance field and the connectivity/neighbourhood for the ORG tree. We adapt all of the three steps to operate upon the intrinsic vertices and edges.
\begin{enumerate}
    \item Firstly, we calculate the geodesic distance using Heat method on the vertices of the intrinsic triangulation. This provides improved robustness to poor triangulations.
    \item Secondly, we work purely within the tangent spaces of each vertex and each face. This requires us to perform parallel transport of gradients onto the same frames of reference whenever we need to do some comparison, such as when we want to identify angle between neighbouring gradients. This affects the final smoothing step too, where we work with barycentric surface points on the faces of the mesh, and transport and interpolate the gradients appropriately.
    \item Finally, when constructing the ORG tree and pruning the ORG tree, we operate purely upon the intrinsic triangulations.  
\end{enumerate}

We find that by working with the intrinsic mesh triangulations, our implementation results in a more robust algorithm. In \cref{fig:comparion-cut-locus-identify} for the mesh \texttt{TID:44395}, we see that our method results in a more stable and consistent cut loci given identical starting points (red), while the original method results in considerably more number of false negatives. For certain meshes, the original method completely fails to work, or perform the homology step of the algorithm~\cite{cutLocusComp}.  Of special importance to \cscd-M is that with our adaptation the target cut locus is the correct one that goes around the geometric feature.

\begin{figure}
    \centering
    \includegraphics[width=0.8\linewidth]{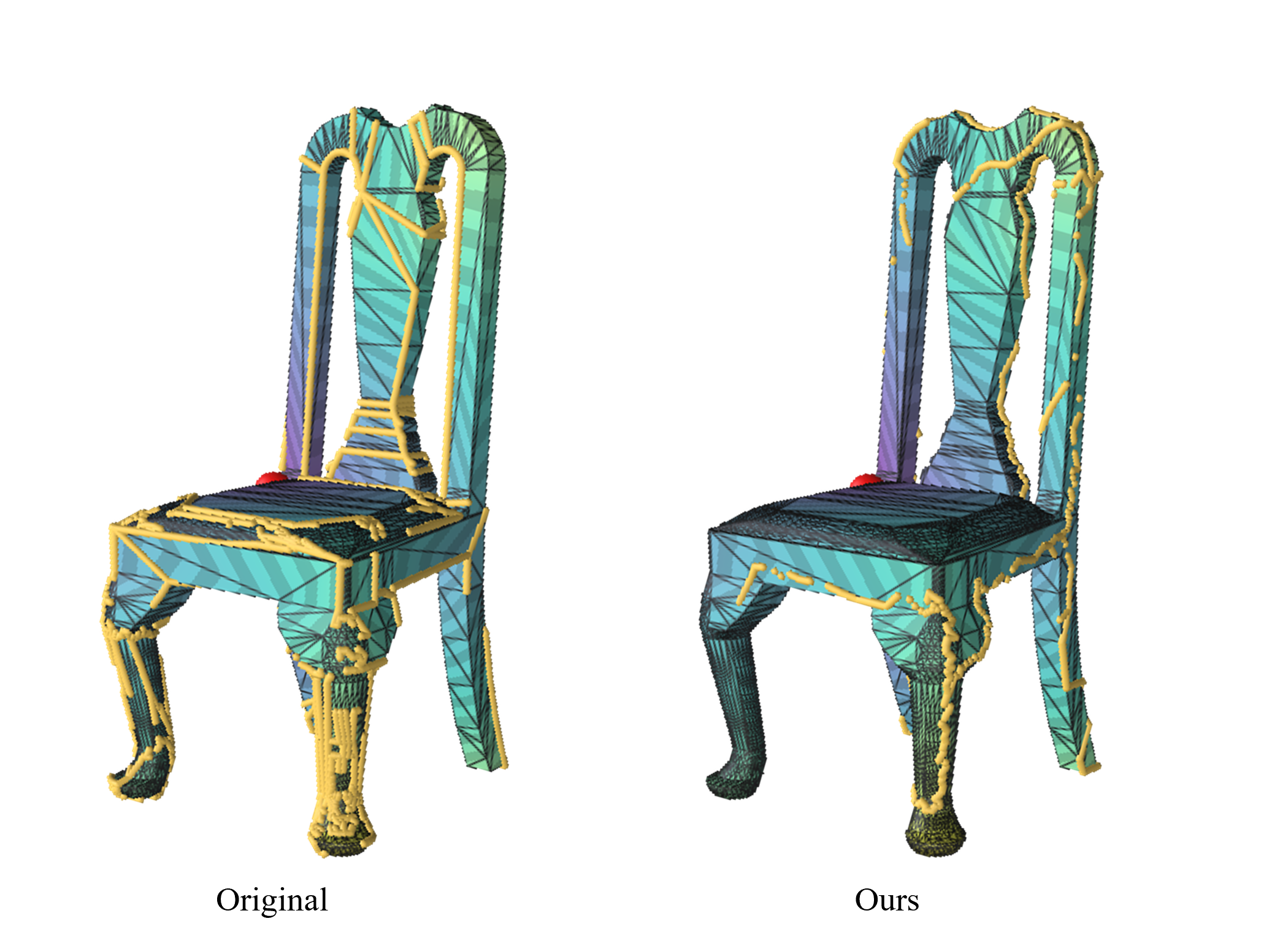}
    \caption{Comparison of our estimated cut loci (right) and the original code (left)~\cite{cutLocusComp}. Furthermore, the original implementation fails to run the homology step of the algorithm. Our adaptation produces a more robust output that selectively chooses the points on the sides away from the source (red) point, while the original code has significant false negatives.}
    \label{fig:comparion-cut-locus-identify}
\end{figure}

\subsection{Our adaptation on point clouds}

We also adapt the practical cut locus algorithm from Ref.~\cite{cutLocusComp} to operate on point clouds. With our adaptation, we have effectively developed a new method for cut locus computation for point clouds. 

Our adaptation resembles the original algorithm. First, we construct a neighbourhood using the $k$ nearest neighbours algorithm. This neighbourhood is then pruned by removing edges that connect points between two previously unconnected regions of the object. We then use the heat method to compute the geodesic distance. We construct the ORG tree associated to the heat distance. The gradients of the distance field are obtained by applying the pre-calculated gradient operator to each point. We apply the remaining process similar to the original algorithm with some changes to account for the irregular structure here. For instance, when calculating the angle deviation the gradient within the neighbourhood of the point, we have to be careful in correctly restricting the search space to the immediate neighbourhood of the vertex.

We illustrate the performance our method on a set of fairly complex objects in \cref{fig:cut-locus-pc}. Our results, though noisy, seem to indicate that the main structure of the cut loci is captured effectively. The noise can be controlled by tuning the free parameters such as the laplacian threshold and the angle deviation threshold of the gradients. For details on those parameters, see Ref.~\cite{cutLocusComp}.

\begin{figure}
    \centering
    \includegraphics[width=0.8\linewidth]{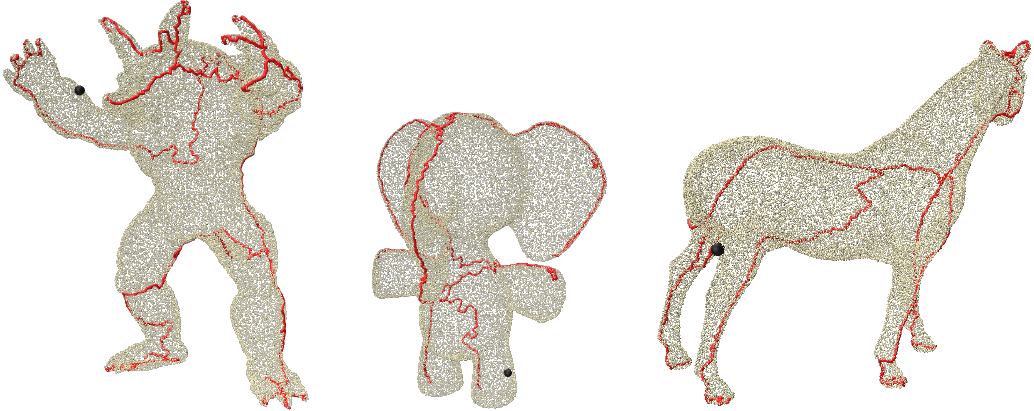}
    \caption{Our adaptation for cut locus identification on point clouds on various point clouds \texttt{armadillo}, \texttt{elephant}, \texttt{horse}. The black point is the source point, and the red curve is the cut locus of the point.}
    \label{fig:cut-locus-pc}
\end{figure}

\section{Additional results}
\label{sec:append-results}

\subsection{Complete Results}

We present the complete quantitative results for our methods and additional methods on a larger set of objects in \cref{tab:quant-result-all}. From the table, we can see that our method performs comparably to both LS~\cite{origLS} and MSLS~\cite{fastLS}; however, MSLS outperforms us consistently. MCF~\cite{tag_mc} performs the worst, but surprisingly outperforms both MSLS and ours on some tubular-shaped objects. As mentioned earlier, there is a large scope for improvement, including developing multiscale approaches. This will help alleviate the primary limitation of our method which is the number of local separators calculated -- by having a multi-scale approach we could scale up the number of separators significantly. 

\begin{table}
    \centering
    \caption{Convolutional Surfaces reconstruction error ($\times 10^{-3}$) for objects using skeletons from \cscd-M (Ours), MCF~\cite{tag_mc}, LS~\cite{origLS} and MSLS~\cite{fastLS}. MCF performs the worst of all the methods. \cscd-M and the two variants of LS are comparable. Though MSLS results in slightly better skeletons. }
    \begin{tabular}{l|c|c|c|c}
    \toprule
       Object  &  LS & MSLS & MCF & \cscd-M \textbf{(Ours)}\\
    \midrule   
        \texttt{armadillo} &  $11.10 $ & $08.27$ & $08.09$ & $08.84$ \\
        \texttt{Copper-key} & $06.13 $ & $04.60$ & $6.02$ & $04.70$ \\
        \texttt{rocker-arm}  & $26.30 $ & $24.40$ & $25.30$ & $24.90$\\
        \texttt{fertility}  & $17.50  $ & $12.20$ & $15.40$ & $13.60$\\
        \texttt{gorilla}  & $6.13$ & $06.71$ & $06.90$ & $07.00 $\\
        \texttt{neptune}  & $04.16$ & $04.82$  & $10.62$ & $03.80$\\
        \texttt{TID:37358}  & $4.852$ & $04.78$  & $11.18$ & $05.08$\\
        \texttt{TID:44395}  & $09.56$ & $10.70$ & $16.80$  & $10.07$\\
        \texttt{TID:39878} & $03.20$ & $03.92$ & $15.83$ & $05.03$ \\
        \texttt{TID:40987} & $09.29$ & $09.81$ & $11.36$ & $08.32$ \\
        \texttt{TID:80516} & $10.40$ & $10.11$ & $09.92$ & $10.49$ \\
        \texttt{TID:82324} & $03.73$ & $04.17$ & $04.17$ & $04.49$ \\
        \texttt{TID:133568} & $04.75$ & $05.49$ & $05.51$ & $04.99$ \\
        \texttt{TID:133079} & $05.27$ & $06.72$ & $05.27$ & $07.03$ \\
        \texttt{Carcinoplax-Suruguensis} & $03.69$ & $03.58$ & $05.47$ & $05.36$ \\
    \midrule
        \textbf{Average} & $08.40$ & $8.02$ & $10.52$ & $08.25$ \\ 
        
    \bottomrule
    \end{tabular}    

    \label{tab:quant-result-all}
\end{table}

\subsection{Results on Complex Meshes}

In \cref{fig:res-complex-shapes}, we also show qualitative results of our method vs LS~\cite{origLS} on a set of complex meshes. We find that our method performs similar to LS when representing the mesh. Here it is set to a fixed value of $N=4096$. In future works, one can explore the automation of this choice, or a better way to choose $N$. 
\begin{figure}
    \centering
    \includegraphics[width=0.95\linewidth]{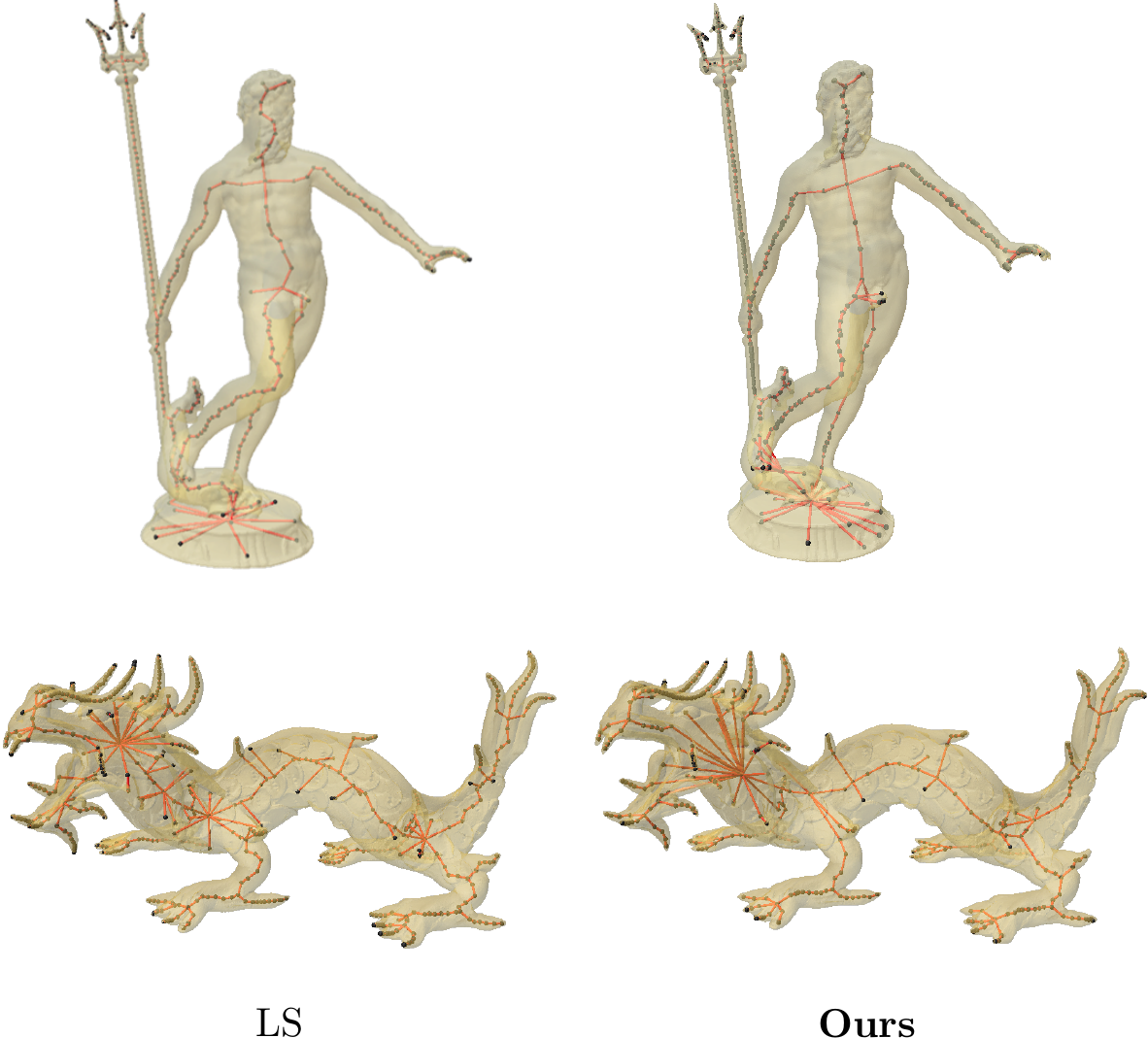}
    \caption{\cscd-M vs LS on complex shapes: Our method results in skeletons that are comparable to LS. Specifically on the \texttt{xyzrgb-dragon} object, our skeleton appears to be smoother and contains fewer noisy branches in the the body of the dragon.}
    \label{fig:res-complex-shapes}
\end{figure}

\begin{figure}
    \centering
    \includegraphics[width=0.6\linewidth]{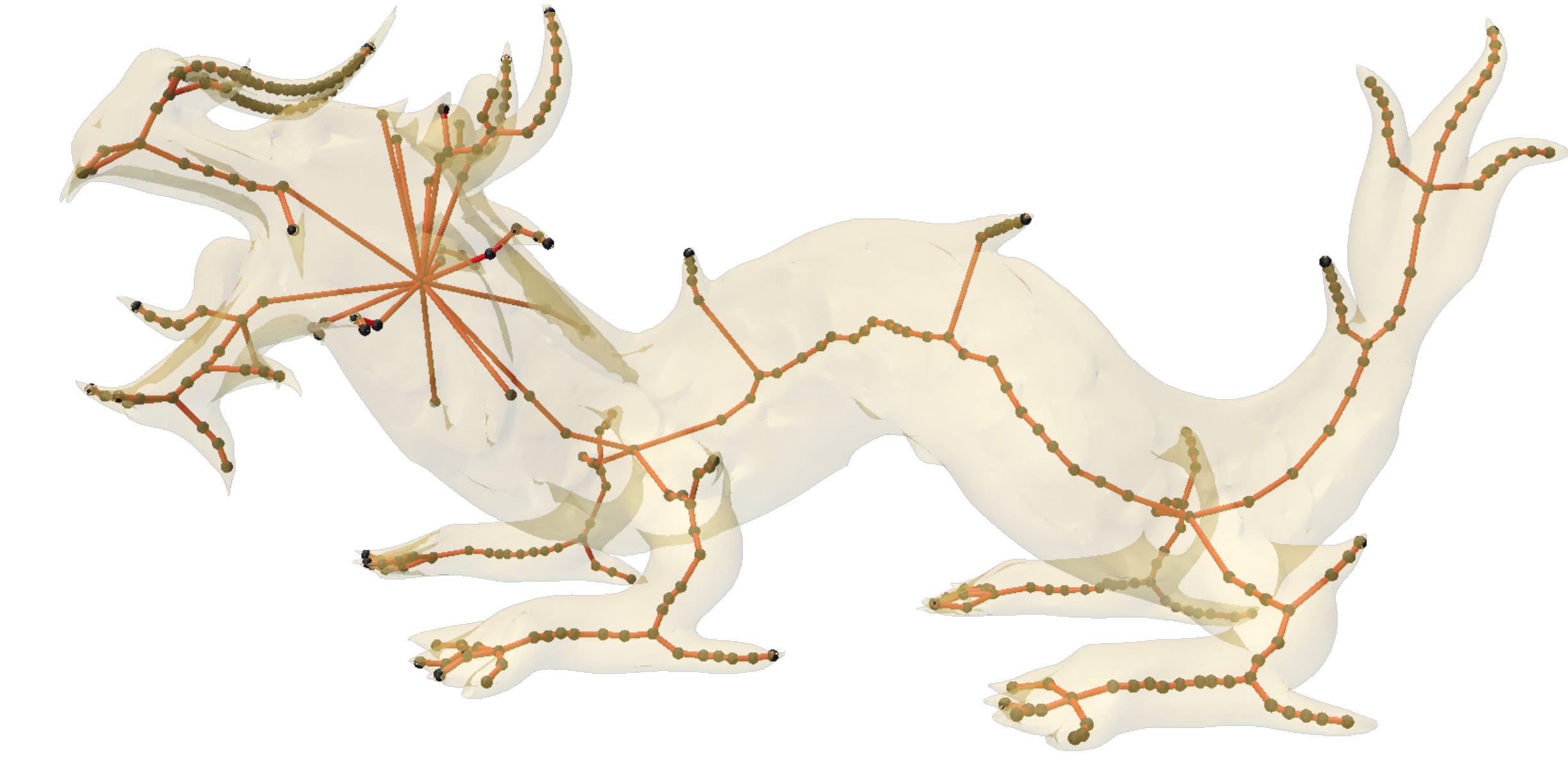}
    \caption{Smoothened \texttt{xyzrgb-dragon} for \cscd-M. We see that the spurious branches have reduced after smoothing the bod}
    \label{fig:xyzrgb-dragon-smooth}
\end{figure}

\subsection{Results on Meshes with Holes}

Our current realization doesn't handle all the cases with holes, as mentioned in Sec~\ref{sec:cscd-desc}.  Nevertheless our framework can in principle handle all these cases. One particular case (namely, with only a single boundary) and its solution is mentioned in Sec.~\ref{sec:cscd-desc}. In that case, we talk about connecting points in the boundary where the gradient seems to point in opposite directions, such that they cancel each other. We, however, note that this would miss few local separators for cases with more than a single boundary. Any two points on two separate boundaries and the path connecting between them could act as local separator too. 

A particular case where our current realization cannot work is when there is a longitudinal hole in a torus such that it may be completely flattened onto a 2D sheet. This is because now the (open) path connecting two points in the two boundaries would act as a local separator.  

In \cref{fig:res-holes}, we show illustrative results for meshes with small holes and how local separators are able to wrap around it. The torus has three holes -- two small holes back-to-back on the right side, and a break on the left side. Notice that our curve skeleton does not close because the torus itself completely breaks at one point. This is topologically valid since the nature of the input object has itself changed. We still outperform both LS and MSLS though. LS completely fails to run on this object with holes and MSLS while producing a curve skeleton does so more poorly compared to our method, changing the topology of the mesh itself.

\begin{figure}
    \centering
    \includegraphics[width=\linewidth]{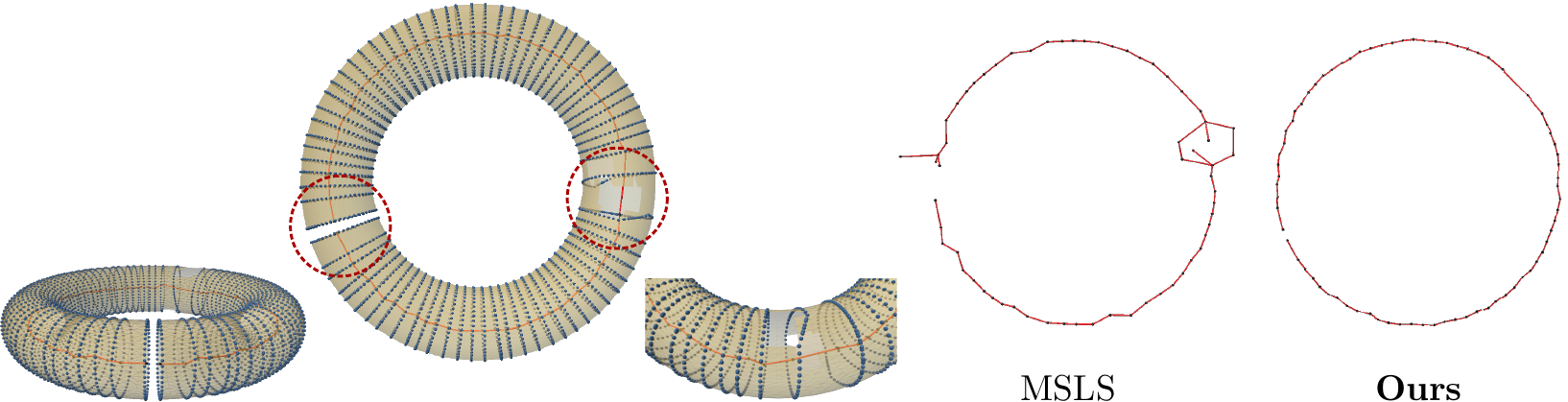}
    \caption{\cscd-M and MSLS on a torus with holes. Left panel, \textbf{\cscd-M local separators on the torus} -- highlighting how the separators go around the holes. Right panel, \textbf{comparison of the curve skeleton} obtained by MSLS and Ours (\cscd-M).}
    \label{fig:res-holes}
\end{figure}

\subsection{Results on Noisy Meshes}
\label{sec:append-noisy-mesh}

We test \cscd-M on a small subset of noisy meshes to evaluate its robustness against noise in the vertex positions. We create the noisy objects by adding standard normal noise proportional to the average edge length $\tau$ of the mesh.

\begin{figure}
    \centering
    \includegraphics[width=0.8\linewidth]{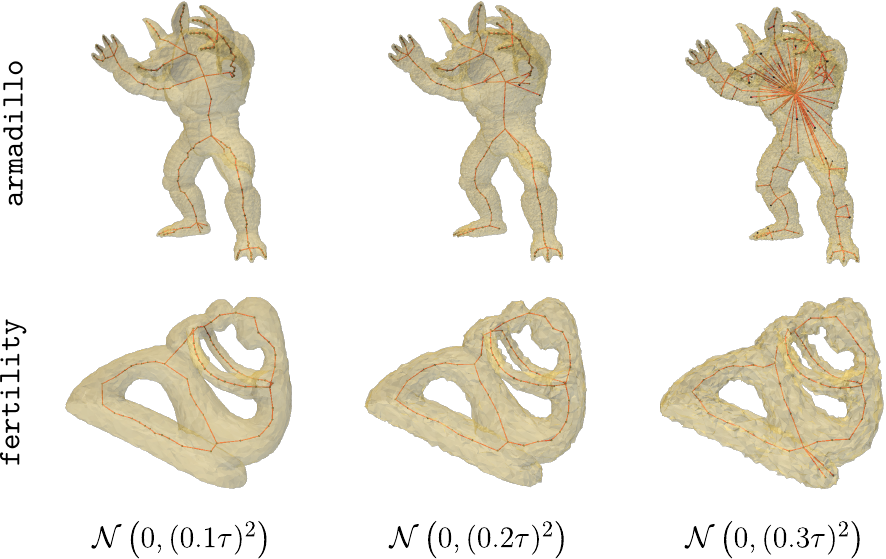}
    \caption{\textbf{Results of \cscd-M on noisy meshes \texttt{armadillo} and \texttt{fertility}:} Here we show the curve skeleton generated at progressively noisier vertex positions. $\mathcal{N}\left(0, (c \tau)^2\right)$ indicates the amount of gaussian noise added to each vertex's position, where $\tau$ is the average edge length of the mesh and $c$ is a constant that controls the amount of noise.}
    \label{fig:res-noisy-mesh}
\end{figure}

\subsection{Results on Mesh Resolution}

We present results with low mesh resolution in Fig.~\ref{fig:res-mesh-resolution}. Both LS and MSLS lead to distorted skeletons especially around the torso of the \texttt{armadillo}. Our method reliably generates the skeleton -- we believe this is due to our ability to capitalize on the dual areas of the vertices which allows us to handle unequal sized faces, thereby creating a better estimate for the 3D positions.

\begin{figure}[htp]
    \centering
    \includegraphics[width=0.5\linewidth]{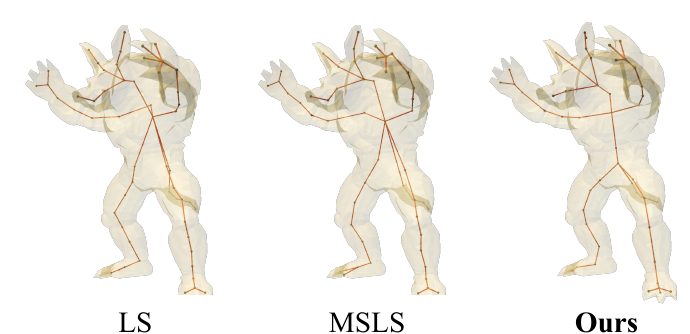}
    \caption{\textbf{Results on a low mesh resolution:} Both LS and MSLS lead to distorted skeletons at reduced mesh resolution while our method still obtains well-centered skeletons. Here the mesh resolution was reduced by $2\times$ the original mesh through mesh decimation.}
    \label{fig:res-mesh-resolution}
\end{figure}

\subsection{Timing Analysis of \cscd-M}
\label{sec:append-inference-times}
We present in \cref{tab:timing-inference}, the inference times for LS, MSLS and \cscd-M on a set of objects. All the inference-time experiments were run on a high-end laptop with a Intel Core i7 processor. In order to make a fair comparison, all methods were restricted to use only a single thread of the CPU.

\noindent\textbf{Why is this in the appendix?}

There are a few reasons as to why we decided to put this analysis within the appendix:
\begin{enumerate}
    \item LS, MSLS and our method use different heuristics for sampling and selecting local separators, so the comparison is not direct. One could argue that LS could, in principle, be sped up by reducing the number of local separators sampled, but that parameter is not available in their codebase; therefore we have to stick to the pre-defined heuristics chosen by the original authors. 
    \item Our method uses a fixed $N=3000$, while the number of separators can vary for both LS and MSLS. We have chosen to implement this simple strategy, but we still observe comparable performance to LS (both qualitatively and quantitatively). We suspect this is due to the differences in the heuristics for sampling, where we use geodesic distance to sample farther away points. Our choice to fix $N$ can go both ways -- on one hand, we could increase $N$ to gain better performance at the cost of inference times, while on the other, we are also overestimating $N$ for simpler objects such as \texttt{fertility}, where $N=1000$ also suffices. In some cases, however, we have taken $N=4000$ on the account of the meshes being more complicated.
    \item As mentioned in the limitations, our realizations \cscd-M and \cscd-PC are simply meant to be a starting point. Our choices in the framework were made due to its simplicity and directness. There are many places (such as the cut locus identification procedure) where one can easily speed up the algorithm. We also believe that our particular codebase can be considerably improved and sped up through better coding practices -- we had to add additional checks and balances due to bugs present in the packages used. One could fix those bugs, or implement them from scratch and yield better speed.
\end{enumerate}

\noindent For the above reasons, we didn't feel it was fair to call this a strong comparison. However, these, still, are inference times of the methods that yield comparable performance (all are within $3\%$ of each other). 

\begin{table}
    \centering
    \caption{Runtime Analysis of \cscd-M (Ours), LS~\cite{origLS} and MSLS~\cite{fastLS} (in seconds). Our results are calculated $N = 3000$, which on average provided similar results to LS. $^*$ However, three cases specifically use $N=4000$ on the account of the mesh being more complex. We report both the average times and the median times since LS seems to have large outliers. $^\dagger$ For LS, we additionally remove \texttt{gorilla} since it takes too long to compute.}
    \begin{tabular}{l|c|c|c|c|c}
        \toprule
        Object & $|V|$ & $|F|$ & LS & MSLS & \cscd-M (Ours)  \\
        \midrule
        \texttt{TID:44395} & $2948$ & $5900$ & $6.781$ & $2.137$ & $37.652$ \\
        \texttt{fertility} & $4494$  & $9000$ & $9.385$ & $2.793$ & $41.434$ \\
        \texttt{TID:37358} & $9780$ & $19560$ & $6.225$ & $3.477$ & $75.91$ \\
        \texttt{Copper\_key} & $10000$ & $20036$ & $13.58$ & $5.485$ & $75.483$ \\
        \texttt{rocker-arm} & $10044$ & $20088$ & $141.032$ & $6.538$ & $75.764$ \\
        \texttt{TID:32770} & $20125$ & $40246$ & $1889.703$ & $13.201$ & $225.1745^*$ \\
        \texttt{gorilla} & $48762$ & $97520$ & $\geq 2400^\dagger$ & $32.458$ & $532.163$ \\
        \texttt{armadillo} & $49990$ & $99976$ & $994.747$ & $33.698$ & $495.326$ \\
        \texttt{garuda-vishnu} & $49972$ & $100084$ & $292.425$ & $30.133$ & $609.304^*$ \\
        \texttt{neptune} & $50000$ & $100008$ & $886.682$ & $31.529$ & $611.6585^*$ \\
        \midrule
        \textbf{Average} & & & 471.1733 & 16.1449 & 277.9869 \\
        \textbf{Median} & & & 141.032 & 9.8695 & 150.54225 \\
        \bottomrule
    \end{tabular}

    \label{tab:timing-inference}
\end{table}

\subsection{Additional Qualitative Results of \cscd-PC}

\begin{figure}[p]
    \centering
    \includegraphics[width=\linewidth]{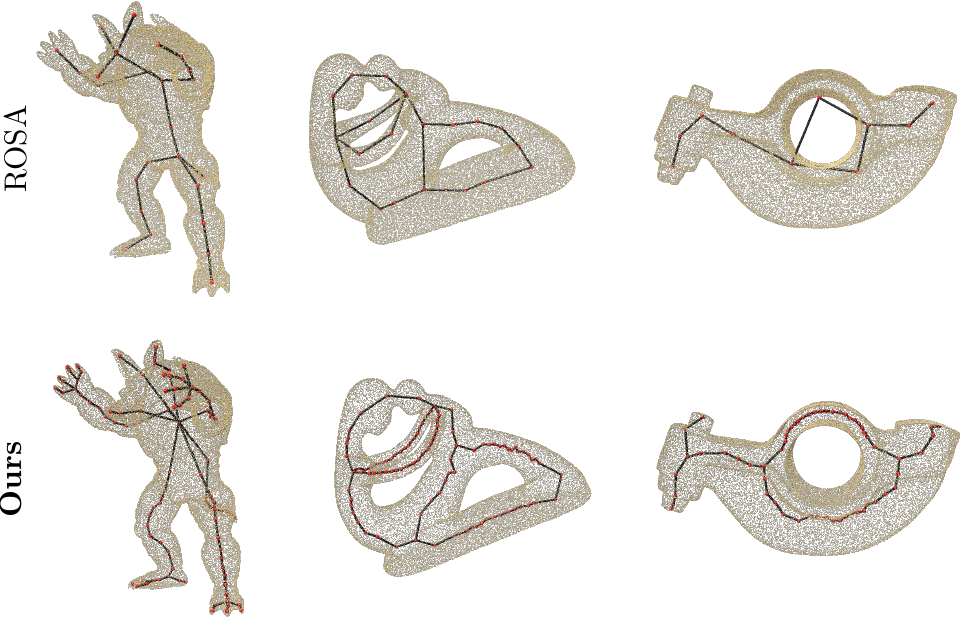}
    \caption{\textbf{\cscd-PC vs ROSA:} Comparative results of our method vs ROSA. We are able to capture the details of the object while ROSA~\cite{rosa} struggles to do so. Specifically, we see that ROSA results in fewer nodes and coarser skeletons compared to \cscd-PC.}
    \label{fig:cscd_pc_vs_rosa}
\end{figure}

In \cref{fig:cscd_pc_vs_rosa}, we show additional qualitative results of \cscd-PC against another popular point cloud skeletonization technique ROSA~\cite{rosa}. Compared to ROSA, our method yields more detailed curve skeletons. Additionally, not shown here, \cscd-PC is faster than ROSA on these objects, but that could simply be an outcome of different programming languages used -- Python vs MATLAB. 

\section{Downstream Applications}
\label{sec:append-downstream-complete}

\subsection{Shape Classification}

We present the details and the results for the shape classification on the Princeton Shape Dataset in Tab.~\ref{tab:shape-classification-append}. We perform the shape classification by using the shape diameter function and comparison function that calculates the 1-D Wasserstein distance between two distributions. With a sufficiently high resolution for the bins of the distribution, 1D wasserstein distance would be robust to the boundaries of the bins.

\begin{table}[htp]
    \centering
    \caption{\textbf{Results on Shape Classification on PSD:} We calculate the Shape Diameter Function (SDF) from the obtained skeletons and use that to classify the a subset of classes from the Princeton Shape Dataset. We find that the SDF obtained from our method works the most reliably at classifying the objects.}
    \begin{tabular}{c|c|c|c}
        \toprule
        Metric & LS & MSLS & \cscd-M \\
        \midrule
        Accuracy & $0.63$ & $0.74$ & $\mathbf{0.79}$ \\ 
        F$1$ Score & $0.59$ & $0.76$ & $\mathbf{0.80}$ \\
        \bottomrule
    \end{tabular}
    \label{tab:shape-classification-append}
\end{table}

\subsection{Shape Segmentation}

We perform unsupervised shape segmentation of various objects in the Princeton Shape Dataset. We show that the segmentation remains consistent over different poses demonstrating the robustness of the skeletonization algorithm to pose changes.

\begin{figure}[htp]
    \centering
    \includegraphics[width=0.5\linewidth]{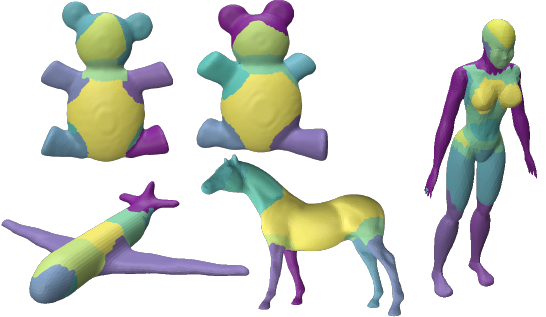}
    \caption{\textbf{[Downstream Application: Shape Segmentation]:} Our method works as a strong skeletonization technique for the shape diameter function (SDF) based unsupervised segmentation. The obtained segmentation is robust to pose variations and works very well.}
    \label{fig:shape-segmentation-append}
\end{figure}

\subsection{Identifying Handles, Tunnels, and Constrictions}
By removing the strong locality constraint during curve shortening, the optimized loop slides toward bottleneck regions. Sampling multiple source points allows us to detect global constrictions, while overlapping loops are pruned. Scoring is then based on loop length. Our framework also automatically identifies handles, pruned in the skeletonization task, tunnels and other constrictions (see Fig.~\ref{fig:tunnel-handle-append}).

\begin{figure}[htp]
    \centering
    \includegraphics[width=0.5\linewidth]{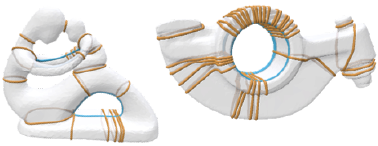}
    \caption{\textbf{[Downstream Application: Handle/Tunnel/Constricting Loop Detection]:} With minor changes, our local separator identification algorithm works effectively to identify handles/tunnels and constricting loops on the shape. These loops form useful basis for other tasks such as surface cutting, etc.}
    \label{fig:tunnel-handle-append}
\end{figure}

\section{Ablations}
\subsection{Choice of Geodesic Distance vs Euclidean Distance}
\label{sec:geodesic-vs-euclid}
We find that while the choice of the geodesic distance results in un-intuitive local separators, the final curve skeletons seem to resemble the one obtained from the Euclidean distance constraints. This could be simply due to the large number of local separators calculated -- eventually leading to similar local separators being calculated for both constraints. 
However there exist many cases (such as one in \cref{fig:res-euclid-geodesic}) where the geodesic based method may fail completely but the Euclidean would work.

In \cref{fig:res-euclid-geodesic}, we illustrate an example where the two methods perform similarly for a wide range of objects. However in the case of \texttt{gorilla}, the Geodesic constraint fails to give a good curve skeleton with missing features and off-axis nodes. This is most likely due to the large abdomen width of the object due to which tracing a path around it is generally longer then going around the nearest arm or leg. However with Euclidean constraints, we are not limited to the surface of the manifold and we can choose the cut loci at the back of the abdomen. See the inset from Sec.~\ref{sec:target-cl} for a visual example of this phenomenon.

\begin{figure}
    \centering
    \includegraphics[height=0.95\textheight]{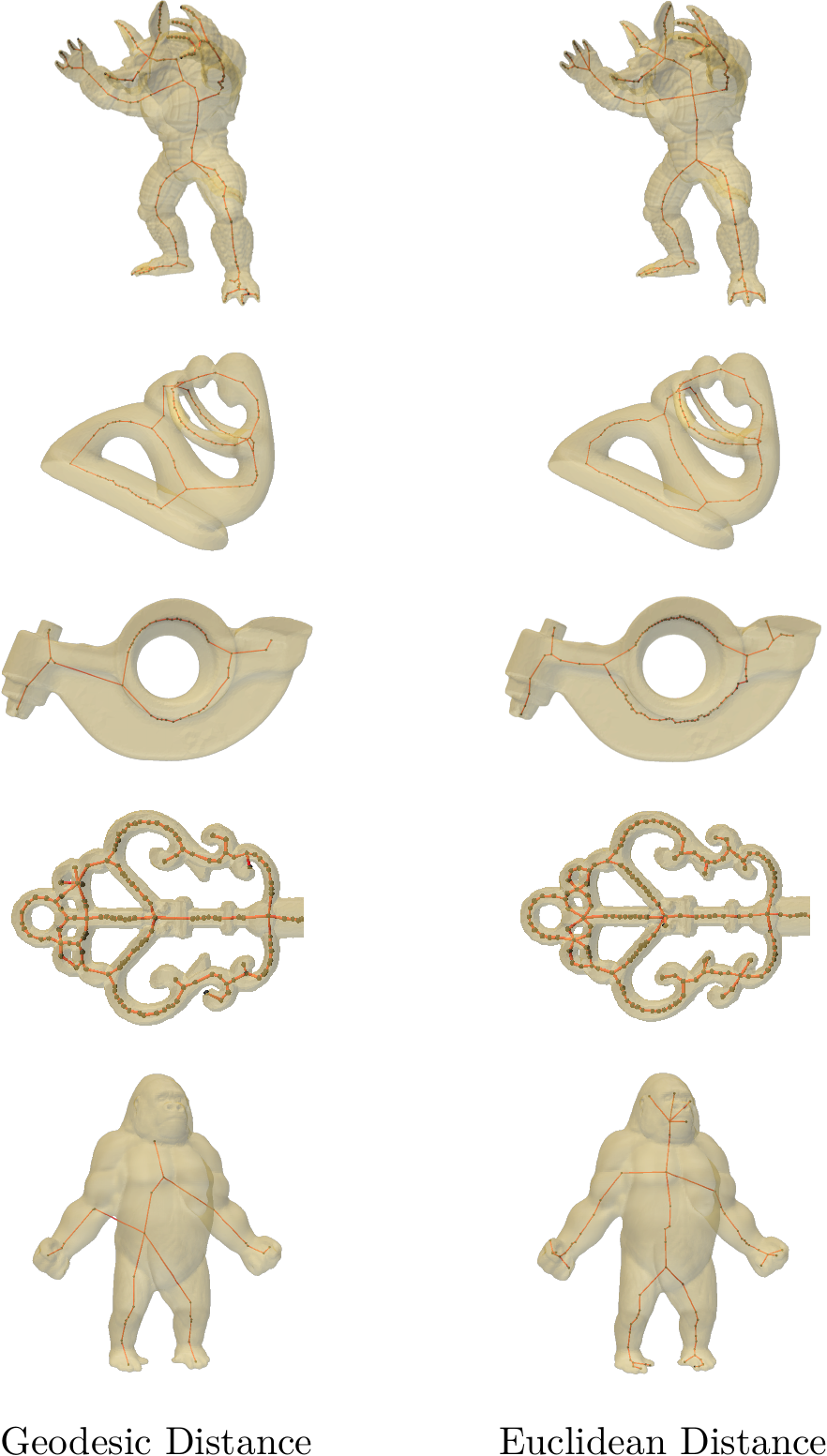}
    \caption{\textbf{Geodesic Constraint vs Euclid Constraint} for calculating the local separators: We see that for many objects the resultant skeleton looks pretty acceptable. However, for certain meshes like \texttt{gorilla}, the geodesic constraint would lead to poor results.}
    \label{fig:res-euclid-geodesic}
\end{figure}

\subsection{Number of Local Separators calculated in Stage 1}
\label{sec:append-ablation-num-ls}

Generally speaking, the number of local separators required is proportional to the complexity of the object in terms of the geometrical features of the mesh. Simpler tubular-like structures like \texttt{fertility} require significantly fewer separators compared to objects like \texttt{armadillo}. In \cref{fig:res-sep-recon} we plot the number of local separators calculated vs the reconstruction error of the object.

\begin{figure}
    \centering
    \includegraphics[width=0.7\linewidth]{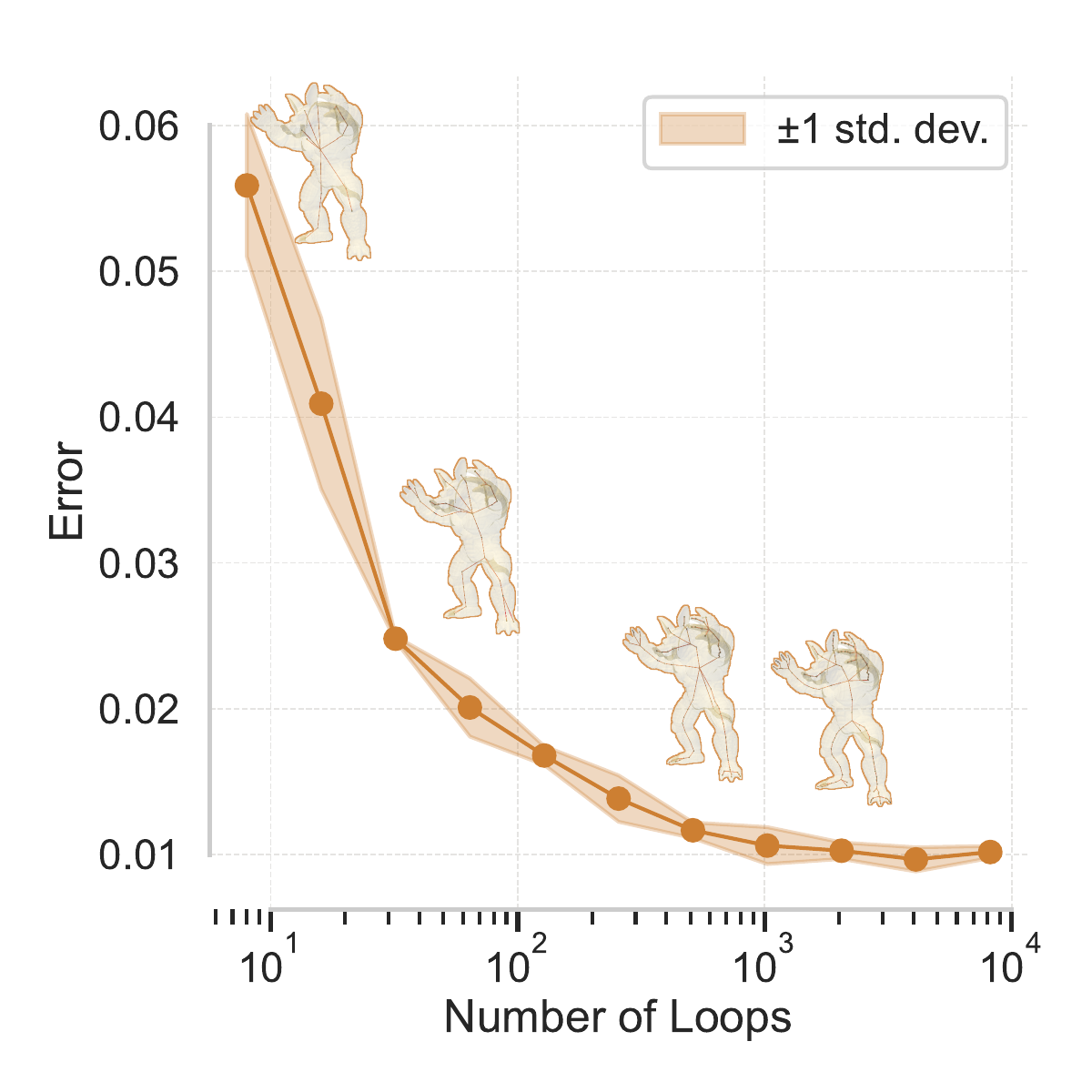}
    \caption{\textbf{Number of local separators vs reconstruction Error for \texttt{armadillo}}: Increasing the number of local separators reduces the error drastically, but the decrease in error varies based on the complexity of the object. The reconstruction errors are evaluated for 8, 16, 32, 64, 128, 256, 512, 1024, 2048, 4096 and 8192 number of local separators.}
    \label{fig:res-sep-recon}
\end{figure}

\begin{figure}
    \centering
    \includegraphics[width=0.7\linewidth]{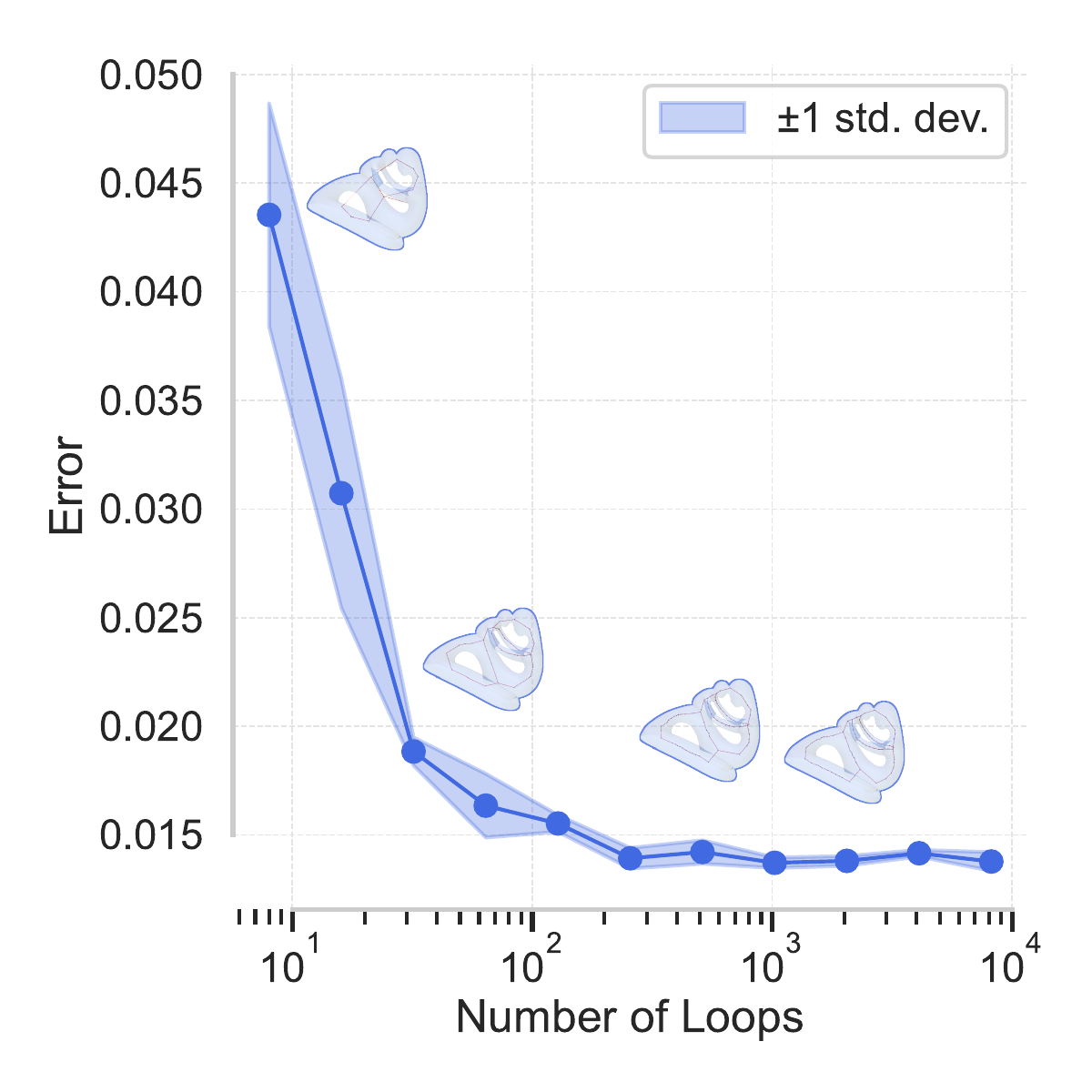}
    \caption{\textbf{Number of local separators vs reconstruction Error for \texttt{fertility}}: Increasing the number of local separators reduces the error drastically, but the decrease in error varies based on the complexity of the object. The reconstruction errors are evaluated for 8, 16, 32, 64, 128, 256, 512, 1024, 2048, 4096 and 8192 number of local separators.}
    \label{fig:res-sep-recon-fertility}
\end{figure}

\begin{figure}
    \centering
    \includegraphics[width=0.7\linewidth]{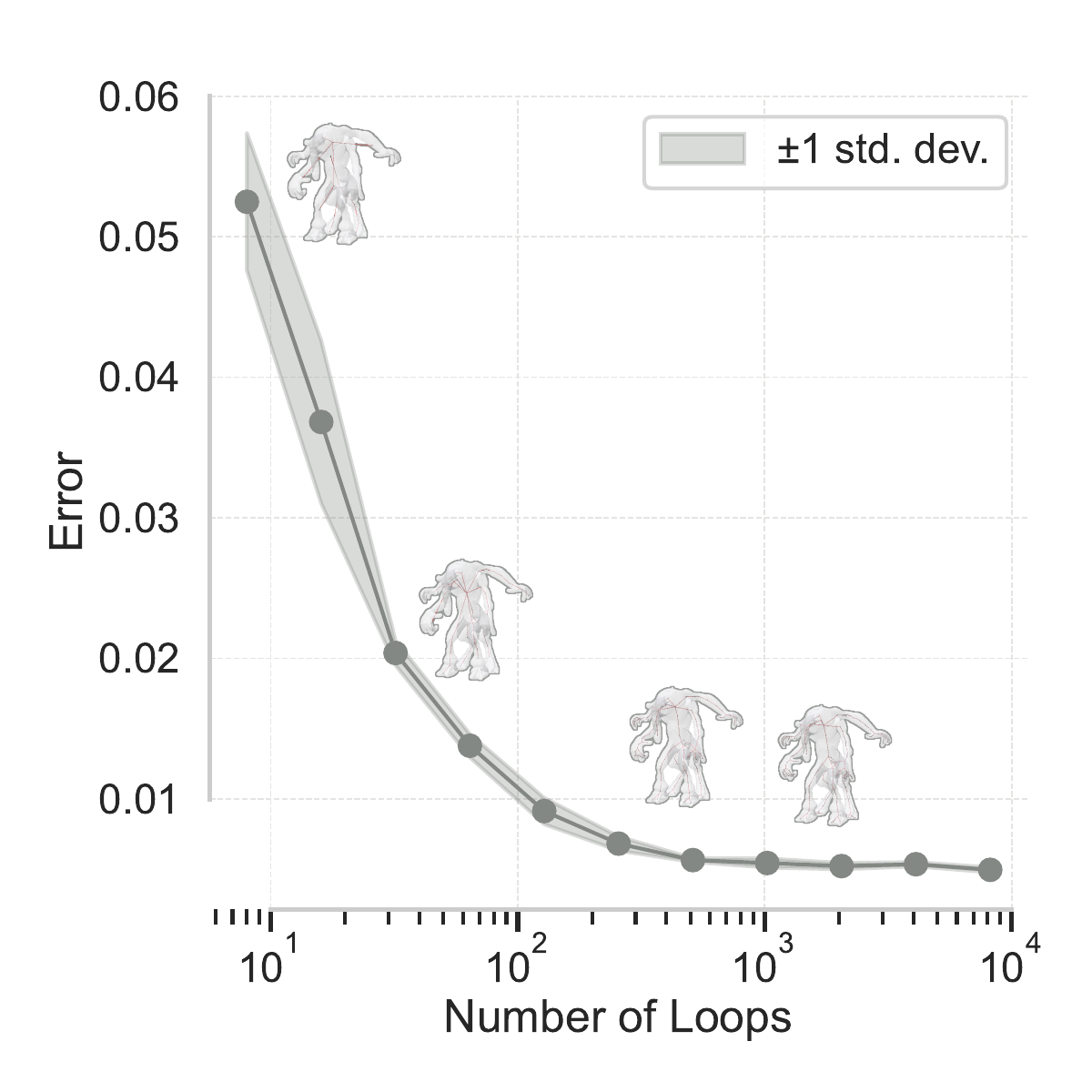}
    \caption{\textbf{Number of local separators vs reconstruction Error for \texttt{TID:133568}}: Increasing the number of local separators reduces the error drastically, but the decrease in error varies based on the complexity of the object. The reconstruction errors are evaluated for 8, 16, 32, 64, 128, 256, 512, 1024, 2048, 4096 and 8192 number of local separators.}
    \label{fig:res-sep-recon-133568}
\end{figure}

\section{Discussions}

We present below a short discussion on some topics that are related to the paper. These discuss some open directions for future work. 

\subsection{On the derivation of LS from \cscd}
\label{sec:append-ls-cscd}

While we would not exactly obtain LS, we would derive a graph version of \cscd that effectively replicates the results from LS. 

\begin{enumerate}
    \item For the geodesic distance, we use a simple graph based distance. 
    \item The target cut locus is then chosen similar to the current method, i.e., based on the minimum Euclidean distance. 
    \item To optimize the curve, we can follow an iterative unfolding scheme, where the path between two vertices is iteratively shortened using Djikstra's shortest path algorithm.
    \item With the optimized local separators, overlap can simply be checked by determining if two local separators share a vertex.
    \item Next, we assign the nearest vertices to each local separator, thereby creating the quotient graph.
    \item Finally, based on the quotient graph, the curve skeleton is constructed and post-processed.
\end{enumerate}

\subsection{On the multiscale version of \cscd-M}

\cscd-M has immense potential for a multi-scale approach. This is because we have the ability to sample points on the face of a low-poly mesh, and operate on these face surface points through barycentric interpolations. In this way, we could, in principle, sample as densely or sparsely as desired -- effectively independent of the polygon count of the mesh. However, the quality of the mesh would influence the accuracy of the quantities being calculated; for example, low-poly meshes could hinder the processes like geodesic distance calculation or edge flip operations. In these cases, one could explore using higher-order corrections to account for these issues. Therefore, unlike other multi-scale approaches, we could get away with directly working upon reduced poly-meshes and introducing higher order corrections, which could make it extremely quick.

\section{Limitations and Future Work}
Our work has a few limitations:
\begin{enumerate}
    \item \cscd requires a defined metric (for the calculation of geodesic distances) don the representation, which may hinder adaptation to certain formats such as NeRFs. However, most common representations --- meshes, point clouds~\cite{lap-nonmanifold}, and digital surfaces~\cite{ddg-discrete-surf} --- provide such metrics. 
    \item Our realizations are intended as starting points, with primary results on meshes and limited tests on point clouds. Future work could enhance individual modules for greater speed, robustness, and performance, and extend the framework to other representations.
\end{enumerate}

\clearpage
\newpage
\section{Terminology and Definitions}
\label{sec:append-vocab}
Here we define some of the technical terms used in the paper:
\begin{description}
  \item[Cut locus / Cut loci] 
    Given a compact Riemannian manifold $(M,g)$ and a point $p\in M$, the \emph{cut locus} of $p$, denoted $\text{Cut}(p)$, is the closure of the set of endpoints of geodesics emanating from $p$ that cease to be globally minimizing beyond those points.  Equivalently, a point $q\in M$ lies in $\text{Cut}(p)$ if there exist two or more distinct minimizing geodesics from $p$ to $q$.  The plural \emph{cut loci} refers to such sets for one or more base points.

  \item[Geodesic] 
    A smooth curve $\gamma\colon I\to M$ on a Riemannian manifold $(M,g)$ is called a \emph{geodesic} if it locally extremizes the arc-length functional.  Equivalently, it satisfies the geodesic equation
    \[
      \nabla_{\dot\gamma}\dot\gamma \;=\; 0,
    \]
    where $\nabla$ is the Levi‐Civita connection associated with $g$.

  \item[Curve shortening] 
    The process of evolving a smooth embedded curve $\gamma(s)\subset M$ under the \emph{curve-shortening flow}
    \[
      \frac{\partial \gamma}{\partial t}
      \;=\;
      k\,\mathbf{n},
    \]
    where $k$ is the (scalar) curvature of the curve and $\mathbf{n}$ its inward unit normal.  This flow decreases the total length of $\gamma$ as quickly as possible at each instant.

  \item[Intrinsic triangulation vs.\ extrinsic triangulation] 
    An \emph{extrinsic triangulation} of a surface $S\subset R^3$ is given by the mesh connectivity and the 3D positions of its vertices, with triangle geometry inherited from the embedding in $R^3$.  An \emph{intrinsic triangulation}, by contrast, is specified solely by edge–length assignments satisfying the triangle inequalities, encoding the surface’s intrinsic metric independently of any embedding.

  \item[Integer coordinate system] 
    An \emph{integer coordinate system} on a mesh is an assignment of integer‐valued coordinates (e.g. in $Z^2$ or $Z^3$) to each vertex, in such a way that all edge‐lengths and combinatorial relationships can be computed exactly using integer arithmetic.  This avoids floating‐point error in geometric algorithms.

  \item[Laplacian / Laplace–Beltrami operator] 
    On a smooth Riemannian manifold $(M,g)$, the \emph{Laplace–Beltrami operator} $\Delta$ acting on a scalar function $f$ is
    \[
      \Delta f \;=\; \div(\nabla f),
    \]
    where $\nabla$ and $\div$ are the gradient and divergence induced by $g$.  On a triangle mesh, the \emph{cotangent Laplacian} at vertex $i$ is discretely given by
    \[
      (\Delta f)_i
      \;=\;
      \frac{1}{2A_i}
      \sum_{j\in N(i)}
        \bigl(\cot\alpha_{ij} + \cot\beta_{ij}\bigr)
        \,(f_j - f_i),
    \]
    where $A_i$ is a local area weight (e.g.\ the Voronoi area) and $\alpha_{ij},\beta_{ij}$ are the angles opposite the edge $(i,j)$.

  \item[Interior Point Method] 
    A class of algorithms for solving constrained optimization problems of the form
    \[
      \min_x\;f(x)
      \quad\text{s.t.}\quad
      g_i(x)\le0,\quad h_j(x)=0,
    \]
    by introducing a barrier term (e.g.\ $-\mu\sum_i\log(-g_i(x))$) for each inequality and solving a sequence of unconstrained problems as the barrier parameter $\mu\to0^+$.  At each iteration, the algorithm stays strictly in the interior of the feasible region and follows a “central path” to the optimum.
\end{description}

\input{sections/2_relWork}

%% file: sections/2_relWork.tex
\section{Related Works}
\label{sec:formatting}

Our work builds upon several fundamental techniques in 3D shape analysis, which we outline in detail below.

\paragraph{Geodesic distance computation on meshes/point clouds:}
Critical to our work are methods for computing geodesic distances on 3D shapes. Geodesic distance computation on meshes has been extensively studied \cite{geodesic_survey}. Of interest to us are PDE-based methods, which assume that the discrete representation is based on a continuous underlying manifold. Heat method \cite{Crane:2017:HMD}, based on the diffusion process, connects Varadhan's formula~\cite{varadhan} to geodesic distances -- thereby reducing the computation to solving a series of linear equations. \cite{Crane:2017:HMD} leads to faster computation with only a small hit to the accuracy. 

\paragraph{Curve Shortening on 3D shapes:} Also critical to our framework are methods for shortening curves on 3D shapes. These methods are often analogous to geodesic computation on surfaces \cite{MARTINEZ2005667,XIN20071081,xin_2011,sharp,LIU2017105,YE201973,yuan_opt}.

\paragraph{Curve Skeletonization for Meshes:} The concept of curve skeletons for 3D objects is not well-defined; this lack of a formal definition has led to the multitude of hand-crafted methods \cite{skel_report, skel_report_2}. Most of these methods rely on the idea that for tubular shapes, there exists a 1D structure that preserves the shape of the topology. These methods are based on the geometrical features of the objects, local decimation of objects, or regional division of objects. Tierny et al. propose an unified method for constructing and simplifying Reeb graphs on 3D meshes using discrete contours to extract affine-invariant, visually meaningful topological skeletons, however, it is not clear how the method performs on pointclouds and other geometric domains \cite{tierny:hal-00725576}. Other representative works include \cite{Au_mesh_contraction,tag_mc,Livesu2012ReconstructingTC,cheng_dual,cov_axis, genus_0,nicu2007curve, cornea2005curve}. \cscd-M is inspired by the LS method\cite{origLS}. MSLS~\cite{fastLS} is the multi-scale variant of LS that allows it to sample more loops efficiently. In principle, constructing a multi-scale variant of \cscd-M should be easy as we can simply work on faces of coarse meshes, but we leave this for a future exploration.

\paragraph{Curve Skeletonization for Point Clouds:} 
Curve skeletonization on point clouds has gained attention recently with increased data availability. Methods like \cite{huang} use medians over centroids, while \cite{rosa} leverages approximate rotational symmetry. However, these approaches often yield lower fidelity and miss fine details. Learning-based methods \cite{p2mat, Lin2020Point2SkeletonLS} struggle to generalize to new meshes. Other recent point cloud methods include EPCS~\cite{LI2023209} and CA++~\cite{dou2022coverage,wang2024coverage}. 

While we have not explicitly evaluated our method on volumetric data, voxel representations can be converted to surface meshes as a preprocessing step prior to skeletonization.\cite{brown2019robust,wickramasinghe2020voxel2mesh,9406392}

%% file: main.bib
@String(CVPR= {IEEE Conf. Comput. Vis. Pattern Recog.})

@String(TOG= {ACM Trans. Graph.})

@String(CVPR  = {CVPR})

@String(TOG   = {ACM TOG})


%% file: sample-base.bib
@String{Computing = "Computing" }

@String{Computer = "{IEEE} Computer" }

@String{Springer = "Springer-Verlag" }

@ARTICLE{9406392,
  author={Lv, Chenlei and Lin, Weisi and Zhao, Baoquan},
  journal={IEEE Transactions on Multimedia}, 
  title={Voxel Structure-Based Mesh Reconstruction From a 3D Point Cloud}, 
  year={2022},
  volume={24},
  number={},
  pages={1815-1829},
  doi={10.1109/TMM.2021.3073265}}

@inproceedings{wickramasinghe2020voxel2mesh,
  title={Voxel2Mesh: 3D mesh model generation from volumetric data},
  author={Wickramasinghe, Udaranga and Remelli, Edoardo and Knott, Graham and Fua, Pascal},
  booktitle={Medical Image Computing and Computer Assisted Intervention--MICCAI 2020: 23rd International Conference, Lima, Peru, October 4--8, 2020, Proceedings, Part IV 23},
  pages={299--308},
  year={2020},
  organization={Springer}
}

@article{brown2019robust,
  title={A Robust Algorithm for Voxel-to-Polygon Mesh Phantom Conversion},
  author={Brown, Justin L and Furuta, Takuya and Bolch, Wesley E},
  journal={Brain and Human Body Modeling: Computational Human Modeling at EMBC 2018},
  pages={317--327},
  year={2019},
  publisher={Springer International Publishing}
}

@inproceedings{cornea2005curve,
  title={Curve-skeleton applications},
  author={Cornea, Nicu D and Silver, Deborah and Min, Patrick},
  booktitle={VIS 05. IEEE Visualization, 2005.},
  pages={95--102},
  year={2005},
  organization={IEEE}
}

@phdthesis{suarez2019modeling,
  title={Modeling shapes with skeletons: scaffolds \& anisotropic convolution},
  author={Suarez, Alvaro Javier Fuentes},
  year={2019},
  school={COMUE Universit{\'e} C{\^o}te d'Azur (2015-2019)}
}

@article{nicu2007curve,
  title={Curve-Skeleton Properties, Applications and Algorithms},
  author={Nicu, D and Silver, C and Silver, D},
  journal={IEEE Transactions on Visualization and Computer Graphics},
  volume={13},
  number={3},
  pages={530--548},
  year={2007}
}

@article{origLS,
author = {B\ae{}rentzen, Andreas and Rotenberg, Eva},
title = {Skeletonization via Local Separators},
year = {2021},
issue_date = {October 2021},
publisher = {Association for Computing Machinery},
address = {New York, NY, USA},
volume = {40},
number = {5},
issn = {0730-0301},
url = {https://doi.org/10.1145/3459233},
doi = {10.1145/3459233},
journal = {ACM Trans. Graph.},
month = {sep},
articleno = {187},
numpages = {18}
}

@inproceedings{tierny:hal-00725576,
  TITLE = {{3D Mesh Skeleton Extraction Using Topological and Geometrical Analyses}},
  AUTHOR = {Tierny, Julien and Vandeborre, Jean-Philippe and Daoudi, Mohamed},
  URL = {https://hal.science/hal-00725576},
  BOOKTITLE = {{14th Pacific Conference on Computer Graphics and Applications (Pacific Graphics 2006)}},
  ADDRESS = {Tapei, Taiwan},
  PAGES = {s1poster},
  YEAR = {2006},
  MONTH = Oct,
  HAL_ID = {hal-00725576},
  HAL_VERSION = {v1},
}

@inproceedings{fastLS,
title = "Multilevel Skeletonization Using Local Separators",
author = "B{\ae}rentzen, {J. Andreas} and Christensen, {Rasmus Emil} and G{\ae}de, {Emil Toftegaard} and Eva Rotenberg",
year = "2023",
doi = "10.4230/LIPIcs.SoCG.2023.13",
language = "English",
volume = "258",
booktitle = "Proceedings of the 39th International Symposium on Computational Geometry (SoCG 2023)"
}

@article{kimmel1998computing,
  title={Computing geodesic paths on manifolds},
  author={Kimmel, Ron and Sethian, James A},
  journal={Proceedings of the national academy of Sciences},
  volume={95},
  number={15},
  pages={8431--8435},
  year={1998},
  publisher={The National Academy of Sciences}
}

@article{surazhsky2005fast,
  title={Fast exact and approximate geodesics on meshes},
  author={Surazhsky, Vitaly and Surazhsky, Tatiana and Kirsanov, Danil and Gortler, Steven J and Hoppe, Hugues},
  journal={ACM transactions on graphics (TOG)},
  volume={24},
  number={3},
  pages={553--560},
  year={2005},
  publisher={Acm New York, NY, USA}
}

@article{Crane:2017:HMD,
 author = {Crane, Keenan and Weischedel, Clarisse and Wardetzky, Max},
 title = {The Heat Method for Distance Computation},
 journal = {Commun. ACM},
 issue_date = {November 2017},
 volume = {60},
 number = {11},
 month = oct,
 year = {2017},
 issn = {0001-0782},
 pages = {90--99},
 numpages = {10},
 url = {http://doi.acm.org/10.1145/3131280},
 doi = {10.1145/3131280},
 acmid = {3131280},
 publisher = {ACM},
 address = {New York, NY, USA},
}

@article{cutLocusComp,
author = {Mancinelli, C. and Livesu, M. and Puppo, E.},
title = {Practical Computation of the Cut Locus on Discrete Surfaces},
journal = {Computer Graphics Forum},
volume = {40},
number = {5},
pages = {261-273},
doi = {https://doi.org/10.1111/cgf.14372},
url = {https://onlinelibrary.wiley.com/doi/abs/10.1111/cgf.14372},
eprint = {https://onlinelibrary.wiley.com/doi/pdf/10.1111/cgf.14372},
year = {2021}
}

@ARTICLE{curveShortening,
  author={Yuan, Na and Wang, Peihui and Meng, Wenlong and Chen, Shuangmin and Xu, Jian and Xin, Shiqing and He, Ying and Wang, Wenping},
  journal={IEEE Transactions on Visualization and Computer Graphics}, 
  title={A Variational Framework for Curve Shortening in Various Geometric Domains}, 
  year={2023},
  volume={29},
  number={4},
  pages={1951-1963},
  doi={10.1109/TVCG.2021.3135021}}

@article{Au_mesh_contraction,
author = {Au, Oscar Kin-Chung and Tai, Chiew-Lan and Chu, Hung-Kuo and Cohen-Or, Daniel and Lee, Tong-Yee},
title = {Skeleton Extraction by Mesh Contraction},
year = {2008},
issue_date = {August 2008},
publisher = {Association for Computing Machinery},
address = {New York, NY, USA},
volume = {27},
number = {3},
issn = {0730-0301},
url = {https://doi.org/10.1145/1360612.1360643},
doi = {10.1145/1360612.1360643},
journal = {ACM Trans. Graph.},
month = {aug},
pages = {1–10},
numpages = {10}
}

@article{li2015q,
  title={Q-mat: Computing medial axis transform by quadratic error minimization},
  author={Li, Pan and Wang, Bin and Sun, Feng and Guo, Xiaohu and Zhang, Caiming and Wang, Wenping},
  journal={ACM Transactions on Graphics (TOG)},
  volume={35},
  number={1},
  pages={1--16},
  year={2015},
  publisher={ACM New York, NY, USA}
}

@article{wang2022computing,
  title={Computing medial axis transform with feature preservation via restricted power diagram},
  author={Wang, Ningna and Wang, Bin and Wang, Wenping and Guo, Xiaohu},
  journal={ACM Transactions on Graphics (TOG)},
  volume={41},
  number={6},
  pages={1--18},
  year={2022},
  publisher={ACM New York, NY, USA}
}

@article{tag_mc,
author = {Tagliasacchi, Andrea and Alhashim, Ibraheem and Olson, Matt and Zhang, Hao},
title = {Mean Curvature Skeletons},
journal = {Computer Graphics Forum},
volume = {31},
number = {5},
pages = {1735-1744},
doi = {https://doi.org/10.1111/j.1467-8659.2012.03178.x},
url = {https://onlinelibrary.wiley.com/doi/abs/10.1111/j.1467-8659.2012.03178.x},
eprint = {https://onlinelibrary.wiley.com/doi/pdf/10.1111/j.1467-8659.2012.03178.x},
year = {2012}
}

@article{Livesu2012ReconstructingTC,
  title={Reconstructing the Curve-Skeletons of 3D Shapes Using the Visual Hull},
  author={Marco Livesu and Fabio Guggeri and Riccardo Scateni},
  journal={IEEE Transactions on Visualization and Computer Graphics},
  year={2012},
  volume={18},
  pages={1891-1901},
  url={https://api.semanticscholar.org/CorpusID:8913957}
}

@article{cheng_dual,
title = {Skeletonization via dual of shape segmentation},
journal = {Computer Aided Geometric Design},
volume = {80},
pages = {101856},
year = {2020},
issn = {0167-8396},
doi = {https://doi.org/10.1016/j.cagd.2020.101856},
url = {https://www.sciencedirect.com/science/article/pii/S0167839620300431},
author = {Jingliang Cheng and Xinyu Zheng and Shuangmin Chen and Guozhu Liu and Shiqing Xin and Lin Lu and Yuanfeng Zhou and Changhe Tu}
}

@article{cov_axis,
author = {Dou, Zhiyang and Lin, Cheng and Xu, Rui and Yang, Lei and Xin, Shiqing and Komura, Taku and Wang, Wenping},
title = {Coverage Axis: Inner Point Selection for 3D Shape Skeletonization},
journal = {Computer Graphics Forum},
volume = {41},
number = {2},
pages = {419-432},
doi = {https://doi.org/10.1111/cgf.14484},
url = {https://onlinelibrary.wiley.com/doi/abs/10.1111/cgf.14484},
eprint = {https://onlinelibrary.wiley.com/doi/pdf/10.1111/cgf.14484},
year = {2022}
}

@ARTICLE{genus_0,
  author={Reniers, Dennie and van Wijk, Jarke and Telea, Alexandru},
  journal={IEEE Transactions on Visualization and Computer Graphics}, 
  title={Computing Multiscale Curve and Surface Skeletons of Genus 0 Shapes Using a Global Importance Measure}, 
  year={2008},
  volume={14},
  number={2},
  pages={355-368},
  doi={10.1109/TVCG.2008.23}
}

@article{huang,
author = {Huang, Hui and Wu, Shihao and Cohen-Or, Daniel and Gong, Minglun and Zhang, Hao and Li, Guiqing and Chen, Baoquan},
title = {L1-Medial Skeleton of Point Cloud},
year = {2013},
issue_date = {July 2013},
publisher = {Association for Computing Machinery},
address = {New York, NY, USA},
volume = {32},
number = {4},
issn = {0730-0301},
url = {https://doi.org/10.1145/2461912.2461913},
doi = {10.1145/2461912.2461913},
journal = {ACM Trans. Graph.},
month = {jul},
articleno = {65},
numpages = {8}
}

@article{rosa,
author = {Tagliasacchi, Andrea and Zhang, Hao and Cohen-Or, Daniel},
title = {Curve Skeleton Extraction from Incomplete Point Cloud},
year = {2009},
issue_date = {August 2009},
publisher = {Association for Computing Machinery},
address = {New York, NY, USA},
volume = {28},
number = {3},
issn = {0730-0301},
url = {https://doi.org/10.1145/1531326.1531377},
doi = {10.1145/1531326.1531377},
journal = {ACM Trans. Graph.},
month = {jul},
articleno = {71},
numpages = {9}
}

@article{p2mat,
title = {P2MAT-NET: Learning medial axis transform from sparse point clouds},
journal = {Computer Aided Geometric Design},
volume = {80},
pages = {101874},
year = {2020},
issn = {0167-8396},
doi = {https://doi.org/10.1016/j.cagd.2020.101874},
url = {https://www.sciencedirect.com/science/article/pii/S0167839620300613},
author = {Baorong Yang and Junfeng Yao and Bin Wang and Jianwei Hu and Yiling Pan and Tianxiang Pan and Wenping Wang and Xiaohu Guo}
}

@article{Lin2020Point2SkeletonLS,
  title={Point2Skeleton: Learning Skeletal Representations from Point Clouds},
  author={Chu-Hsing Lin and Changjian Li and Yuan Liu and Nenglun Chen and Yi-King Choi and Wenping Wang},
  journal={2021 IEEE/CVF Conference on Computer Vision and Pattern Recognition (CVPR)},
  year={2020},
  pages={4275-4284},
  url={https://api.semanticscholar.org/CorpusID:227238925}
}

@article{varadhan,
author = {Varadhan, S. R. S.},
title = {On the behavior of the fundamental solution of the heat equation with variable coefficients},
journal = {Communications on Pure and Applied Mathematics},
volume = {20},
number = {2},
pages = {431-455},
doi = {https://doi.org/10.1002/cpa.3160200210},
url = {https://onlinelibrary.wiley.com/doi/abs/10.1002/cpa.3160200210},
eprint = {https://onlinelibrary.wiley.com/doi/pdf/10.1002/cpa.3160200210},
year = {1967}
}

@article{MARTINEZ2005667,
title = {Computing geodesics on triangular meshes},
journal = {Computers and Graphics},
volume = {29},
number = {5},
pages = {667-675},
year = {2005},
issn = {0097-8493},
doi = {https://doi.org/10.1016/j.cag.2005.08.003},
url = {https://www.sciencedirect.com/science/article/pii/S0097849305001299},
author = {Dimas Martínez and Luiz Velho and Paulo C. Carvalho}
}

@article{XIN20071081,
title = {Efficiently determining a locally exact shortest path on polyhedral surfaces},
journal = {Computer-Aided Design},
volume = {39},
number = {12},
pages = {1081-1090},
year = {2007},
issn = {0010-4485},
doi = {https://doi.org/10.1016/j.cad.2007.08.001},
url = {https://www.sciencedirect.com/science/article/pii/S0010448507001959},
author = {Shi-Qing Xin and Guo-Jin Wang}
}

@ARTICLE{xin_2011,
  author={Xin, Shi-Qing and He, Ying and Fu, Chi-Wing},
  journal={IEEE Transactions on Visualization and Computer Graphics}, 
  title={Efficiently Computing Exact Geodesic Loops within Finite Steps}, 
  year={2012},
  volume={18},
  number={6},
  pages={879-889},
  doi={10.1109/TVCG.2011.119}
}

@article{sharp,
author = {Sharp, Nicholas and Crane, Keenan},
title = {You Can Find Geodesic Paths in Triangle Meshes by Just Flipping Edges},
year = {2020},
issue_date = {December 2020},
publisher = {Association for Computing Machinery},
address = {New York, NY, USA},
volume = {39},
number = {6},
issn = {0730-0301},
url = {https://doi.org/10.1145/3414685.3417839},
doi = {10.1145/3414685.3417839},
journal = {ACM Trans. Graph.},
month = {nov},
articleno = {249},
numpages = {15}
}

@article{gillespie2021integer,
  title={Integer Coordinates for Intrinsic Geometry Processing},
  author={Gillespie, Mark and Sharp, Nicholas and Crane, Keenan},
  journal={arXiv preprint arXiv:2106.00220},
  year={2021}
}

@article{LIU2017105,
title = {An optimization-driven approach for computing geodesic paths on triangle meshes},
journal = {Computer-Aided Design},
volume = {90},
pages = {105-112},
year = {2017},
note = {SI:SPM2017},
issn = {0010-4485},
doi = {https://doi.org/10.1016/j.cad.2017.05.022},
url = {https://www.sciencedirect.com/science/article/pii/S0010448517301033},
author = {Bangquan Liu and Shuangmin Chen and Shi-Qing Xin and Ying He and Zhen Liu and Jieyu Zhao}
}

@article{YE201973,
title = {DE-Path: A Differential-Evolution-Based Method for Computing Energy-Minimizing Paths on Surfaces},
journal = {Computer-Aided Design},
volume = {114},
pages = {73-81},
year = {2019},
issn = {0010-4485},
doi = {https://doi.org/10.1016/j.cad.2019.05.025},
url = {https://www.sciencedirect.com/science/article/pii/S0010448519302015},
author = {Zipeng Ye and Yong-Jin Liu and Jianmin Zheng and Kai Hormann and Ying He}
}

@ARTICLE {yuan_opt,
author = {N. Yuan and P. Wang and W. Meng and S. Chen and J. Xu and S. Xin and Y. He and W. Wang},
journal = {IEEE Transactions on Visualization and; Computer Graphics},
title = {A Variational Framework for Curve Shortening in Various Geometric Domains},
year = {2023},
volume = {29},
number = {04},
issn = {1941-0506},
pages = {1951-1963},
doi = {10.1109/TVCG.2021.3135021},
publisher = {IEEE Computer Society},
address = {Los Alamitos, CA, USA},
month = {apr}
}

@article{Kordalewski2013NewGH,
  title={New Greedy Heuristics For Set Cover and Set Packing},
  author={David Kordalewski},
  journal={ArXiv},
  year={2013},
  volume={abs/1305.3584},
  url={https://api.semanticscholar.org/CorpusID:16936500}
}

@ARTICLE{distance_fields,
  author={Jones, M.W. and Baerentzen, J.A. and Sramek, M.},
  journal={IEEE Transactions on Visualization and Computer Graphics}, 
  title={3D distance fields: a survey of techniques and applications}, 
  year={2006},
  volume={12},
  number={4},
  pages={581-599},
  doi={10.1109/TVCG.2006.56}}

@article{nerf,
  title={NeRF: Representing Scenes as Neural Radiance Fields for View Synthesis},
  author={Ben Mildenhall and Pratul P. Srinivasan and Matthew Tancik and Jonathan T. Barron and Ravi Ramamoorthi and Ren Ng},
  journal={Commun. ACM},
  year={2020},
  volume={65},
  pages={99-106},
  url={https://api.semanticscholar.org/CorpusID:213175590}
}

@article{nerf_review,
  title={Nerf: Neural radiance field in 3d vision, a comprehensive review},
  author={Gao, Kyle and Gao, Yina and He, Hongjie and Lu, Dening and Xu, Linlin and Li, Jonathan},
  journal={arXiv preprint arXiv:2210.00379},
  year={2022}
}

@inproceedings {mesh_proessing,
booktitle = {Eurographics 2000 - STARs},
editor = {},
title = {{Geometric Signal Processing on Polygonal Meshes}},
author = {Taubin, G.},
year = {2000},
publisher = {Eurographics Association},
ISSN = {1017-4656},
DOI = {10.2312/egst.20001029}
}

@article{deep_learning_mesh,
author = {Wang, He and Zhang, Juyong},
year = {2022},
month = {02},
pages = {},
title = {A Survey of Deep Learning-Based Mesh Processing},
volume = {10},
journal = {Communications in Mathematics and Statistics},
doi = {10.1007/s40304-021-00246-7}
}

@article{skel_report,
author = {Tagliasacchi, Andrea and Delame, Thomas and Spagnuolo, Michela and Amenta, Nina and Telea, Alexandru},
title = {3D Skeletons: A State-of-the-Art Report},
journal = {Computer Graphics Forum},
volume = {35},
number = {2},
pages = {573-597},
doi = {https://doi.org/10.1111/cgf.12865},
url = {https://onlinelibrary.wiley.com/doi/abs/10.1111/cgf.12865},
eprint = {https://onlinelibrary.wiley.com/doi/pdf/10.1111/cgf.12865},
year = {2016}
}

@article{skel_report_2,
title = {A survey on skeletonization algorithms and their applications},
journal = {Pattern Recognition Letters},
volume = {76},
pages = {3-12},
year = {2016},
note = {Special Issue on Skeletonization and its Application},
issn = {0167-8655},
doi = {https://doi.org/10.1016/j.patrec.2015.04.006},
url = {https://www.sciencedirect.com/science/article/pii/S0167865515001233},
author = {Punam K. Saha and Gunilla Borgefors and Gabriella {Sanniti di Baja}}
}

@INPROCEEDINGS{skel_shape_match,
  author={Sundar, H. and Silver, D. and Gagvani, N. and Dickinson, S.},
  booktitle={2003 Shape Modeling International.}, 
  title={Skeleton based shape matching and retrieval}, 
  year={2003},
  volume={},
  number={},
  pages={130-139},
  doi={10.1109/SMI.2003.1199609}}

@InProceedings{transfer4d,
    author    = {Maheshwari, Shubh and Narain, Rahul and Hebbalaguppe, Ramya},
    title     = {Transfer4D: A Framework for Frugal Motion Capture and Deformation Transfer},
    booktitle = {Proceedings of the IEEE/CVF Conference on Computer Vision and Pattern Recognition (CVPR)},
    month     = {June},
    year      = {2023},
    pages     = {12836-12846}
}

@INPROCEEDINGS{skel_recon,
  author={Durix, Bastein and Morin, Géraldine and Chambon, Sylvie and Roudet, Céline and Garnier, Lionel},
  booktitle={2015 International Conference on 3D Vision}, 
  title={Towards Skeleton Based Reconstruction: From Projective Skeletonization to Canal Surface Estimation}, 
  year={2015},
  volume={},
  number={},
  pages={545-553},
  doi={10.1109/3DV.2015.67}}

@article{geodesic_survey,
  title={A Survey of Algorithms for Geodesic Paths and Distances},
  author={Keenan Crane and Marco Livesu and Enrico Puppo and Yipeng Qin},
  journal={ArXiv},
  year={2020},
  volume={abs/2007.10430},
  url={https://api.semanticscholar.org/CorpusID:220665672}
}

@article{Fleishman2005RobustML,
  title={Robust moving least-squares fitting with sharp features},
  author={Shachar Fleishman and Daniel Cohen-Or and Cl{\'a}udio T. Silva},
  journal={ACM SIGGRAPH 2005 Papers},
  year={2005},
  url={https://api.semanticscholar.org/CorpusID:473208}
}

@misc{libigl,
  title = { {libigl}: A simple {C++} geometry processing library},
  author = {Alec Jacobson and Daniele Panozzo and others},
  note = {https://libigl.github.io/},
  year = {2018},
}

@article{geometrycentral,
  title={GeometryCentral: A modern C++ library of data structures and algorithms for geometry processing},
  author={Nicholas Sharp and Keenan Crane and others},
  howpublished="\url{https://geometry-central.net/}",
  year={2019},
  journal="github"
}

@article{open3d,
    author    = {Qian-Yi Zhou and Jaesik Park and Vladlen Koltun},
    title     = {{Open3D}: {A} Modern Library for {3D} Data Processing},
    journal   = {arXiv:1801.09847},
    year      = {2018},
}

@article{scikit-learn,
 title={Scikit-learn: Machine Learning in {P}ython},
 author={Pedregosa, F. and Varoquaux, G. and Gramfort, A. and Michel, V.
         and Thirion, B. and Grisel, O. and Blondel, M. and Prettenhofer, P.
         and Weiss, R. and Dubourg, V. and Vanderplas, J. and Passos, A. and
         Cournapeau, D. and Brucher, M. and Perrot, M. and Duchesnay, E.},
 journal={Journal of Machine Learning Research},
 volume={12},
 pages={2825--2830},
 year={2011}
}

@article{GOH2008326,
title = {Strategies for shape matching using skeletons},
journal = {Computer Vision and Image Understanding},
volume = {110},
number = {3},
pages = {326-345},
year = {2008},
note = {Similarity Matching in Computer Vision and Multimedia},
issn = {1077-3142},
doi = {https://doi.org/10.1016/j.cviu.2007.09.013},
url = {https://www.sciencedirect.com/science/article/pii/S107731420700149X},
author = {Wooi-Boon Goh}
}

@article{lap-nonmanifold,
author = {Sharp, Nicholas and Crane, Keenan},
title = {A Laplacian for Nonmanifold Triangle Meshes},
journal = {Computer Graphics Forum},
volume = {39},
number = {5},
pages = {69-80},
doi = {https://doi.org/10.1111/cgf.14069},
url = {https://onlinelibrary.wiley.com/doi/abs/10.1111/cgf.14069},
eprint = {https://onlinelibrary.wiley.com/doi/pdf/10.1111/cgf.14069},
year = {2020}
}

@InProceedings{ddg-discrete-surf,
author="Coeurjolly, David
and Lachaud, Jacques-Olivier",
editor="Baudrier, {\'E}tienne
and Naegel, Beno{\^i}t
and Kr{\"a}henb{\"u}hl, Adrien
and Tajine, Mohamed",
title="A Simple Discrete Calculus for Digital Surfaces",
booktitle="Discrete Geometry and Mathematical Morphology",
year="2022",
publisher="Springer International Publishing",
address="Cham",
pages="341--353",
isbn="978-3-031-19897-7"
}

@inproceedings{shilane2004princeton,
  title={The princeton shape benchmark},
  author={Shilane, Philip and Min, Patrick and Kazhdan, Michael and Funkhouser, Thomas},
  booktitle={Proceedings Shape Modeling Applications, 2004.},
  pages={167--178},
  year={2004},
  organization={IEEE}
}

@article{LI2023209,
title = {EPCS: Endpoint-based part-aware curve skeleton extraction for low-quality point clouds},
journal = {Computers \& Graphics},
volume = {117},
pages = {209-221},
year = {2023},
issn = {0097-8493},
doi = {https://doi.org/10.1016/j.cag.2023.10.023},
url = {https://www.sciencedirect.com/science/article/pii/S0097849323002601},
author = {Chunhui Li and Mingquan Zhou and Guohua Geng and Yifei Xie and Yuhe Zhang and Yangyang Liu},
}

@inproceedings{wang2024coverage,
  title={Coverage Axis++: Efficient Inner Point Selection for 3D Shape Skeletonization},
  author={Wang, Zimeng and Dou, Zhiyang and Xu, Rui and Lin, Cheng and Liu, Yuan and Long, Xiaoxiao and Xin, Shiqing and Komura, Taku and Yuan, Xiaoming and Wang, Wenping},
  booktitle={Computer Graphics Forum},
  volume={43},
  number={5},
  pages={e15143},
  year={2024},
  organization={Wiley Online Library}
}

@inproceedings{dou2022coverage,
  title={Coverage axis: Inner point selection for 3d shape skeletonization},
  author={Dou, Zhiyang and Lin, Cheng and Xu, Rui and Yang, Lei and Xin, Shiqing and Komura, Taku and Wang, Wenping},
  booktitle={Computer Graphics Forum},
  volume={41},
  number={2},
  pages={419--432},
  year={2022},
  organization={Wiley Online Library}
}

@article{chu2023robustly,
  author = {Chu, Y. and Wang, W. and Li, L.},
  title = {Robustly Extracting Concise 3D Curve Skeletons by Enhancing the Capture of Prominent Features},
  journal = {IEEE Trans Vis Comput Graph},
  volume = {29},
  number = {8},
  pages = {3472--3488},
  year = {2023},
  month = {Aug},
  doi = {10.1109/TVCG.2022.3161962},
  note = {Epub 2023 Jun 29, PMID: 35324442}
}

@article{shapira2008consistent,
  title={Consistent mesh partitioning and skeletonisation using the shape diameter function},
  author={Shapira, Lior and Shamir, Ariel and Cohen-Or, Daniel},
  journal={The Visual Computer},
  volume={24},
  pages={249--259},
  year={2008},
  publisher={Springer}
}

@ArtifactSoftware{R,
    title = {R: A Language and Environment for Statistical Computing},
    author = {{R Core Team}},
    organization = {R Foundation for Statistical Computing},
    address = {Vienna, Austria},
    year = {2019},
    url = {https://www.R-project.org/},
}
